\documentclass{emulateapj}
\usepackage{times,mathptm}
\usepackage{lscape}

\usepackage{epsf}
\usepackage{psfig}
\usepackage{lscape}

\def\etal{{et\,al.\,}}

\def\ltsima{$\; \buildrel < \over \sim \;$}
\def\simlt{\lower.5ex\hbox{\ltsima}}
\def\gtsima{$\; \buildrel > \over \sim \;$}
\def\simgt{\lower.5ex\hbox{\gtsima}}

\begin{document}

\title{The NuSTAR Extragalactic Survey:\\ A First Sensitive Look at the High-Energy Cosmic X-ray Background Population}


\author{D.~M. Alexander,\altaffilmark{1} 
D. Stern,\altaffilmark{2}
A. Del Moro,\altaffilmark{1} 
G.~B. Lansbury,\altaffilmark{1} 
R.~J. Assef,\altaffilmark{2,3}
J. Aird,\altaffilmark{1}
M. Ajello,\altaffilmark{4}
D.~R. Ballantyne,\altaffilmark{5}
F.~E. Bauer,\altaffilmark{6,7}
S.~E. Boggs,\altaffilmark{4}
W.~N. Brandt,\altaffilmark{8,9}
F.~E. Christensen,\altaffilmark{10}
F. Civano,\altaffilmark{11,12}
A. Comastri,\altaffilmark{13}
W.~W. Craig,\altaffilmark{10,14}
M. Elvis,\altaffilmark{12}
B.~W. Grefenstette,\altaffilmark{15}
C.~J. Hailey,\altaffilmark{16}
F.~A. Harrison,\altaffilmark{15}
R.~C. Hickox,\altaffilmark{11}
B. Luo,\altaffilmark{8,9}
K.~K. Madsen,\altaffilmark{15}
J.~R. Mullaney,\altaffilmark{1}
M. Perri,\altaffilmark{17,18}
S. Puccetti,\altaffilmark{17,18}
C. Saez,\altaffilmark{6}
E. Treister,\altaffilmark{19}
C.~M. Urry,\altaffilmark{20}
W.~W. Zhang\altaffilmark{21}
C.~R. Bridge,\altaffilmark{22}
P.~R.~M. Eisenhardt,\altaffilmark{2}
A.~H. Gonzalez,\altaffilmark{23}
S.~H. Miller,\altaffilmark{22} and
C.~W. Tsai\altaffilmark{2}
}

\affil{$^1$~Department of Physics, Durham University, Durham DH1 3LE,
  UK}
\affil{$^2$~Jet Propulsion Laboratory, California Institute of
  Technology, 4800 Oak Grove Drive, Mail Stop 169-221, Pasadena, CA
  91109, USA}
\affil{$^3$~NASA Postdoctoral Program Fellow}
\affil{$^4$~Space Sciences Laboratory, University of California,
  Berkeley, CA 94720, USA}
\affil{$^5$~Center for Relativistic Astrophysics, School of Physics,
  Georgia Institute of Technology, Atlanta, GA 30332, USA}
\affil{$^6$~Pontificia Universidad Cat\'{o}lica de Chile, Departamento
  de Astronom\'{i}a y Astrof\'{i}sica, Casilla 306, Santiago 22,
  Chile}
\affil{$^7$~Space Science Institute, 4750 Walnut Street, Suite 205,
  Boulder, CO 80301, USA}
\affil{$^8$~Department of Astronomy \& Astrophysics, 525 Davey Lab,
  The Pennsylvania State University, University Park, PA 16802, USA}
\affil{$^9$~Institute for Gravitation and the Cosmos, The
  Pennsylvania State University, University Park, PA 16802, USA}
\affil{$^{10}$~DTU Space--National Space Institute, Technical
  University of Denmark, Elektrovej 327, 2800 Lyngby, Denmark}
\affil{$^{11}$~Department of Physics and Astronomy, Dartmouth College,
  6127 Wilder Laboratory, Hanover, NH 03755, USA}
\affil{$^{12}$~Harvard-Smithsonian Center for Astrophysics, 60 Garden
  Street, Cambridge, MA 02138, USA}
\affil{$^{13}$~INAF - Osservatorio Astronomico di Bologna, Via Ranzani
  1, I-40127 Bologna, Italy}
\affil{$^{14}$~Lawrence Livermore National Laboratory, Livermore,
  California 94550, USA}
\affil{$^{15}$~Cahill Center for Astrophysics, 1216 East California
  Boulevard, California Institute of Technology, Pasadena, CA 91125,
  USA}
\affil{$^{16}$~Columbia Astrophysics Laboratory, 550 W 120th Street,
  Columbia University, NY 10027, USA}
\affil{$^{17}$~ASI - Science Data Center, via Galileo Galilei, 00044,
  Frascati, Italy}
\affil{$^{18}$~INAF - Osservatorio Astronomico di Roma, via Frascati
  33, 00040, Monteporzio Catone, Italy}
\affil{$^{19}$~Universidad de Concepci\'{o}n, Departamento de
  Astronom\'{\i}a, Casilla 160-C, Concepci\'{o}n, Chile}
\affil{$^{20}$~Yale Center for Astronomy \& Astrophysics, Yale
  University, Physics Department, PO Box 208120, New Haven, CT
  06520-8120, USA}
\affil{$^{21}$~NASA Goddard Space Flight Center, Greenbelt, Maryland
  20771, USA}
\affil{$^{22}$~California Institute of Technology, MS249-17, Pasadena,
  CA 91125, USA}
\affil{$^{23}$~Department of Astronomy, University of Florida,
  Gainesville, FL 32611-2055, USA}

\shorttitle{THE NUSTAR SERENDIPITOUS SURVEY}

\shortauthors{ALEXANDER ET AL.}

%
\begin{abstract} 
%

We report on the first ten identifications of sources serendipitously
detected by the {\it Nuclear Spectroscopic Telescope Array ({\it
    NuSTAR})} to provide the first sensitive census of the cosmic
X-ray background (CXB) source population at $\simgt10$~keV. We find
that these {\it NuSTAR}-detected sources are $\approx$~100 times
fainter than those previously detected at $\simgt10$~keV and have a
broad range in redshift and luminosity ($z=$~0.020--2.923 and $L_{\rm
  10-40 keV}\approx4\times10^{41}$--$5\times10^{45}$~erg~s$^{-1}$);
the median redshift and luminosity are $z\approx$~0.7 and $L_{\rm
  10-40 keV}\approx3\times10^{44}$~erg~s$^{-1}$, respectively. We
characterize these sources on the basis of broad-band
$\approx$~0.5--32~keV spectroscopy, optical spectroscopy, and
broad-band ultraviolet-to-mid-infrared SED analyzes. We find that the
dominant source population is quasars with $L_{\rm 10-40
  keV}>10^{44}$~erg~s$^{-1}$, of which $\approx$~50\% are obscured
with $N_{\rm H}\simgt10^{22}$~cm$^{-2}$. However, none of the ten {\it
  NuSTAR} sources are Compton thick ($N_{\rm
  H}\simgt10^{24}$~cm$^{-2}$) and we place a 90\% confidence upper
limit on the fraction of Compton-thick quasars ($L_{\rm 10-40
  keV}>10^{44}$~erg~s$^{-1}$) selected at $\simgt10$~keV of
$\simlt33$\% over the redshift range $z=$~0.5--1.1.  We jointly fitted
the rest-frame $\approx$~10--40~keV data for all of the non-beamed
sources with $L_{\rm 10-40 keV}>10^{43}$~erg~s$^{-1}$ to constrain the
average strength of reflection; we find $R<1.4$ for $\Gamma=1.8$,
broadly consistent with that found for local AGNs observed at
$\simgt10$~keV. We also constrain the host galaxy masses and find a
median stellar mass of $\approx10^{11}$~$M_{\odot}$, a factor
$\approx$~5 times higher than the median stellar mass of nearby
high-energy selected AGNs, which may be at least partially driven by
the order of magnitude higher X-ray luminosities of the {\it NuSTAR}
sources. Within the low source-statistic limitations of our study, our
results suggest that the overall properties of the {\it NuSTAR}
sources are broadly similar to those of nearby high-energy selected
AGNs but scaled up in luminosity and mass.

\end{abstract}

\keywords{galaxies: active --- galaxies: high-redshift --- infrared:
  galaxies --- X-rays}


%
\section{Introduction}
%

The cosmic X-ray background (CXB) was first discovered in the early
1960's (Giacconi et~al. 1962), several years before the detection of
the cosmic microwave background (CMB; Penzias \& Wilson
1965). However, unlike the CMB, which is truly diffuse in origin, the
CXB is dominated by the emission from high-energy distant point
sources: Active Galactic Nuclei (AGNs), the sites of intense
black-hole growth that reside at the centers of galaxies (see Brandt
\& Hasinger 2005; Brandt \& Alexander 2010 for reviews). A key goal of
high-energy astrophysics is to determine the detailed composition of
the CXB in order to understand the evolution of AGNs.

Huge strides in revealing the composition of the CXB have been made
over the past decade, with sensitive surveys undertaken by the {\it
  Chandra} and {\it XMM-Newton} observatories (e.g.,\ Alexander
et~al. 2003a; Hasinger et~al. 2007; Brunner et~al. 2008; Luo
et~al. 2008; Comastri et~al. 2011; Xue et~al. 2011). These surveys are
so deep that they have resolved $\approx$~70--90\% of the CXB at
energies of $\approx$~0.5--8~keV (e.g.,\ Worsley et~al. 2005; Hickox
\& Markevitch 2006; Lehmer et~al. 2012; Xue et~al. 2012), revealing a
plethora of obscured and unobscured AGNs out to
$z\approx$~5--6. However, although revolutionary, {\it Chandra} and
{\it XMM-Newton} are only sensitive to sources detected at
$\approx$~0.5--10~keV, far from the peak of the CXB at
$\approx$~20--30~keV (e.g.,\ Frontera et~al. 2007; Ajello et~al. 2008;
Moretti et~al. 2009; Ballantyne et~al. 2011). Until recently, the most
powerful observatories with sensitivity at $\approx$~20--30~keV have
only resolved $\approx$~1--2\% of the CXB at these energies
(e.g.,\ Krivonos et~al. 2007; Ajello et~al. 2008, 2012; Bottacini
et~al. 2012) and therefore provide a limited view of the dominant
source populations (e.g.,\ Sazonov \& Revnivtsev 2004; Markwardt
et~al. 2005; Bassani et~al. 2006; Treister, Urry, \& Virani 2009; Bird
et~al. 2010; Tueller et~al. 2010; Burlon et~al. 2011).

A great breakthrough in resolving the peak of the CXB is the {\it
  Nuclear Spectroscopic Telescope Array (NuSTAR)} observatory. {\it
  NuSTAR} was successfully launched on June 13th 2012 and is the first
$>10$~keV orbiting observatory with focusing optics (Harrison
et~al. 2013). {\it NuSTAR}'s focusing optics provide a $\approx$~1
order of magnitude improvement in angular resolution and a $\approx$~2
orders of magnitude improvement in sensitivity over
previous-generation $>10$~keV observatories, a revolutionary leap
forward in performance. One of the primary objectives of {\it NuSTAR}
is to complete a sensitive extragalactic survey and identify the
source populations that produce the peak of the CXB.

The {\it NuSTAR} extragalactic survey is comprised of three components
(see Table~6 of Harrison et~al. 2013): a deep small-area survey in the
Extended {\it Chandra} Deep Field-South (E-CDF-S; Lehmer et~al. 2005)
field, a medium wider-area survey in the Cosmic Evolution Survey
(COSMOS; Scoville et~al. 2007) field, and a large area (typically
shallow) serendipitous survey conducted in the fields of other {\it
  NuSTAR} targets, including $\approx$~100 {\it Swift}-BAT identified
AGNs. In this paper we report on the first ten spectroscopically
identified sources in the {\it NuSTAR} serendipitous survey. In \S2 we
present the {\it NuSTAR} observations of the serendipitous sources,
the multi-wavelength data, and the details of our data processing
approaches, in \S3 we describe our analysis of the X-ray and
multi-wavelength data, in \S4 we present our results, and in \S5 we
outline our conclusions. We adopt $H_{0}=71$~km~s$^{-1}$~Mpc$^{-1}$,
$\Omega_{M}=0.27$ and $\Omega_{\Lambda}=0.73$ throughout.

%
\section{NuSTAR Observations and Multi-Wavelength Data}
%

{\it NuSTAR} is the first high-energy ($>10$~keV) orbiting observatory
with focusing optics and has a usable energy range of 3--79~keV
(Harrison et~al. 2013). {\it NuSTAR} consists of two co-aligned X-ray
telescopes (focal length of 10.14~m) which focus X-ray photons onto
two independent shielded focal plane modules (FPMs), referred to here
as FPMA and FPMB. Each FPM consists of 4 CdZnTe chips and has a
$\approx12^{\prime}\times12^{\prime}$ field of view at 10~keV; the
pixel size is $2{\farcs}46$. The focusing optics provide {\it NuSTAR}
with a $\approx$~1 order of magnitude improvement in angular
resolution over previous observatories at $>10$~keV; the full-width
half maximum (FWHM) of the point-spread function (PSF) is
$\approx18^{\prime\prime}$ and the half-power diameter is
$\approx58^{\prime\prime}$. The absolute astrometric accuracy of {\it
  NuSTAR} is $\pm5^{\prime\prime}$ (90\% confidence) for bright X-ray
sources and the spectral resolution is $\approx$~0.4~keV (FWHM) at
10~keV.

\subsection{The NuSTAR serendipitous survey}

The {\it NuSTAR} serendipitous survey is the largest-area component of
the {\it NuSTAR} extragalactic survey programme. The serendipitous
survey is built up from {\it NuSTAR}-detected sources in the fields of
{\it NuSTAR} targets, similar in principle to the serendipitous
surveys undertaken in the fields of {\it Chandra} and {\it XMM-Newton}
sources (e.g.,\ Harrison et~al. 2003; Kim et~al. 2004; Watson
et~al. 2009). A major component of the {\it NuSTAR} serendipitous
survey are $\approx$~15--20~ks observations of $\approx$~100 {\it
  Swift}-BAT identified AGNs, which provide both high-quality
high-energy constraints of local AGNs and $\approx$~2--3~deg$^2$ of
areal coverage to search for serendipitous sources. However, the
serendipitous survey is not restricted to these fields and the {\it
  NuSTAR} observations of targets not in the E-CDF-S, COSMOS, and
Galactic-plane surveys are used to search for serendipitous {\it
  NuSTAR} sources; the exposures for these targets are also often
substantially deeper than the {\it NuSTAR} observations of the {\it
  Swift}-BAT AGNs (up-to on-axis exposures of 177.1~ks in the current
paper). The expected areal coverage of the {\it NuSTAR} serendipitous
survey in the first two years is $\approx$~3--4~deg$^{2}$.

Using the {\it NuSTAR} data processing and source detection approach
outlined below, at the time of writing we have serendipitously
detected $\approx$~50 sources in the fields of $\approx$~70 {\it
  NuSTAR} targets. Here we present the properties of the first ten
spectroscopically identified sources; see Table~1. These ten sources
were selected from {\it NuSTAR} observations taken up until January
31st 2013. The selection of these sources for spectroscopic follow-up
observations was based on their visibility to ground-based telescopes
and they should therefore be representative of the overall high-energy
source population.

\subsubsection{Data processing and source searching}

The Level 1 data products were processed with the {\it NuSTAR} Data
Analysis Software (NuSTARDAS) package (v. 0.9.0). Event files (level 2
data products) were produced, calibrated, and cleaned using standard
filtering criteria with the {\tt nupipeline} task and the latest
calibration files available in the {\it NuSTAR} CALDB. The {\it
  NuSTAR} observations of the Geminga field were comprised of 15
separate exposures, which we combined using {\sc ximage}
v4.5.1;\footnote{See
  http://heasarc.gsfc.nasa.gov/docs/xanadu/ximage/ximage.html for
  details of {\sc ximage}.} the other {\it NuSTAR} observations
reported here were individual exposures.

We produced 3--24~keV, 3--8~keV, and 8--24~keV images using {\sc
  dmcopy} from the {\it Chandra} Interactive Analysis Observations
({\sc ciao}) software (v4.4; Fruscione et~al. 2006) for both {\it
  NuSTAR} FPMs.\footnote{See http://cxc.harvard.edu/ciao/index.html
  for details of {\sc ciao}.} We also produced exposure maps in each
energy band for both FPMs, which take account of the fall in the
effective area of the mirrors with off-axis angle and are normalised
to the effective exposure of a source located at the aim point.

We searched for serendipitous sources in all of the six images
(i.e.,\ the three energy bands for each FPM) using {\sc wavdetect}
(Freeman et~al. 2002) with an initial false-positive probability
threshold of $10^{-6}$ and wavelet scales of 4, 5.66, 8, 11.31, and 16
pixels. To be considered a reliable {\it NuSTAR} source we require a
detection to satisfy at least one of two criteria: (1) to be detected
in at least one of the three images for both FPMA and FPMB or (2) to
be detected in at least one of the three images in a single FPM but to
have a lower-energy X-ray counterpart (e.g.,\ detected by {\it
  Chandra}, {\it Swift}-XRT, or {\it XMM-Newton}). Following \S3.4.1
of Alexander et~al. (2003a), we also ran {\sc wavdetect} at a
false-positive probability threshold of $10^{-4}$ to search the six
images (i.e.,\ the three energy bands for each FPM) for lower
significance counterparts of sources already detected at a
false-positive probability threshold of $10^{-6}$ in any of the three
energy bands.

See Tables 1--2 for the details of the X-ray data for the first 10
spectroscopically identified serendipitous {\it NuSTAR} sources. All
of the {\it NuSTAR} sources are detected at $>8$~keV in at least one
FPM.

\subsubsection{Source photometry}

We measured the number of counts for each source at 3--24, 3--8, and
8--24~keV using either a 30$^{\prime\prime}$, 45$^{\prime\prime}$, or
60$^{\prime\prime}$ radius circular aperture centered on the 3--24~keV
{\sc wavdetect} position for each FPM; the encircled energy fractions
of these apertures are $\approx$~0.50, $\approx$~0.66, and
$\approx$~0.77 of the full PSF, respectively, for a source at the aim
point. The choice of aperture is dictated by the brightness of the
source and how close it lies to another source; see Table~2 for the
adopted aperture of each source. These measurements provide the gross
source counts, which we correct for background counts to provide the
net source counts. To obtain a good sampling of the background counts
while minimising the contribution to the background from the source
counts, we measured the background in source-free regions using at
least four circular apertures of 45$^{\prime\prime}$ or
60$^{\prime\prime}$ radius at least 90$^{\prime\prime}$ from the
source. The gross source counts are corrected for the background
counts to give the net source counts, rescaling for the different
sizes of the source and background regions. Errors on the net source
counts are determined as the square root of the gross source
counts. Upper limits are calculated when a source is not detected in
one of the six images or if the net counts are less than the
1~$\sigma$ uncertainty; 3~$\sigma$ upper limits are calculated as 3
times the square root of the gross source counts. See Table~2 for the
source photometry.

\subsubsection{Source fluxes}

The source fluxes are calculated using the net count rates (i.e.,\ the
net counts divided by the source exposure time) and the measured X-ray
spectral slope, following a procedure analogous to that used in the
{\it Chandra} deep field surveys (e.g.,\ Brandt et~al. 2001; Alexander
et~al. 2003a). The X-ray spectral slope is determined from the band
ratio, which we define here as the 8--24~keV/3--8~keV count-rate
ratio. To convert the band ratio into an X-ray spectral slope we used
{\sc xspec} v12.7.1d (Arnaud 1996) and the Response Matrix File (RMF)
and Ancillary Response File (ARF) of the detected {\it NuSTAR}
sources; we produced the RMF and ARF following \S2.1.5. We also used
{\sc xspec} and the RMF and ARF to determine the relationship between
count rate, X-ray spectral slope, and source flux in each of the three
energy bands: 3--24~keV, 3--8~keV, and 8--24~keV. We calculated the
source fluxes in the three energy bands using the observed count rate
and the derived X-ray spectral slope; for the faint {\it NuSTAR}
sources with $<100$ net counts summed over the two FPMs, we set the
X-ray spectral slope to $\Gamma=1.8$, consistent with the average
X-ray spectral slope of the overall sample (see \S4.3). The source
fluxes in each band were then corrected to the 100\% encircled-energy
fraction of the PSF and averaged over the two FPMs.

\subsubsection{Source positions}

To provide the most accurate {\it NuSTAR} source positions and assist
in source matching, we calculated a counts-weighted source
position. This is determined from the 3--24~keV net counts and the
3--24~keV source position in each FPM. If a source is only detected in
one FPM at 3--24~keV then the position of the source in that FPM is
used.\footnote{We derive the {\it NuSTAR} source name from the
  counts-weighted {\it NuSTAR} source position, adjusted to an
  appropriate level of precision (based on the {\it NuSTAR} positional
  accuracy), using the International Astronomical Union (IAU) approved
  naming convention for {\it NuSTAR} sources:
  NuSTAR~JHHMMSS$\pm$DDMM.m, where m is the truncated fraction of an
  arcmin in declination for the arcseconds component.}

\subsubsection{Extraction of the X-ray spectral products}

We extracted the {\it NuSTAR} data to be used in the X-ray spectral
fitting analyzes. The {\it NuSTAR} data were extracted using the {\it
  NuSTAR}-developed software {\tt nuproducts}. {\tt nuproducts}
extracts source and background spectra and produces the RMF and ARF
required to fit the X-ray data; the source and background spectra were
extracted from each FPM using the same-sized apertures and regions as
those adopted for the source photometry.

For the serendipitous source in the Geminga field
(NuSTAR~J063358+1742.4) we combined the source and background spectrum
from each of the 15 observations (see \S2.1.1) to produce a total
source and background spectrum. We also produced an average ARF file
for NuSTAR~J063358+1742.4 by combining the individual ARF files,
weighted by the exposure time for each ARF, and we used the RMF
produced from the first observation when fitting the X-ray data.

\subsection{Lower-energy X-ray data}

To extend the X-ray spectral fitting constraints and assist in the
identification of optical counterparts, we searched for $<10$~keV
counterparts for each {\it NuSTAR}-detected source using {\it
  Chandra}, {\it Swift}-XRT, and {\it XMM-Newton} observations. Since
the {\it NuSTAR} serendipitous programme targets fields containing
well-known Galactic and extragalactic targets, they all have
lower-energy X-ray coverage. However, the only lower-energy X-ray data
available in the IC~751 field is a short ($\approx$~2.3~ks) {\it
  Swift}-XRT observation in which the serendipitous {\it NuSTAR}
source is detected with only 10~counts by XRT, which is insufficient
to provide useful $<10$~keV constraints. For all of the other {\it
  NuSTAR} sources there are good-quality $<10$~keV data and, in some
cases, there was more than one observation available. When selecting
suitable lower-energy data we preferentially chose contemporaneous
observations (i.e.,\ observations taken within $\approx$~1 week of the
{\it NuSTAR} observations), which was the case for three sources in
our sample (NuSTAR~J032459-0256.1; NuSTAR~J121027+3929.1;
NuSTAR~J183443+3237.8). In the absence of contemporaneous observations
we used existing lower-energy data where the 3--8~keV flux agreed to
within a factor of two of the 3--8~keV flux measured from the {\it
  NuSTAR} data; see \S2.3 for more details.

%
%
\begin{figure}[t]
\centerline{\includegraphics[angle=0,width=8.0cm]{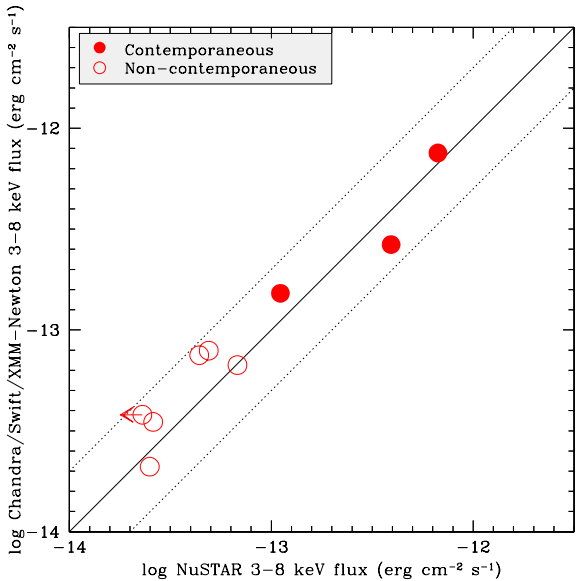}}
\caption{Comparison of the measured fluxes in the 3--8~keV band
  between {\it NuSTAR} and lower-energy X-ray observations ({\it
    Chandra}, {\it Swift}-XRT, or {\it XMM-Newton}); see Table~1 for
  details of the data used for each source. The filled circles
  indicate sources where the lower-energy observations were obtained
  $<1$~week of the date of the {\it NuSTAR} observations and the
  unfilled circles indicate sources where the lower-energy
  observations were obtained $>1$~week of the date of the {\it NuSTAR}
  observations. The solid line indicates agreement between the fluxes
  while the dotted lines indicate a factor of two disagreement between
  the fluxes.}
\end{figure}

%
%
\begin{figure*}[t]
\centerline{\includegraphics[angle=0,width=15.0cm]{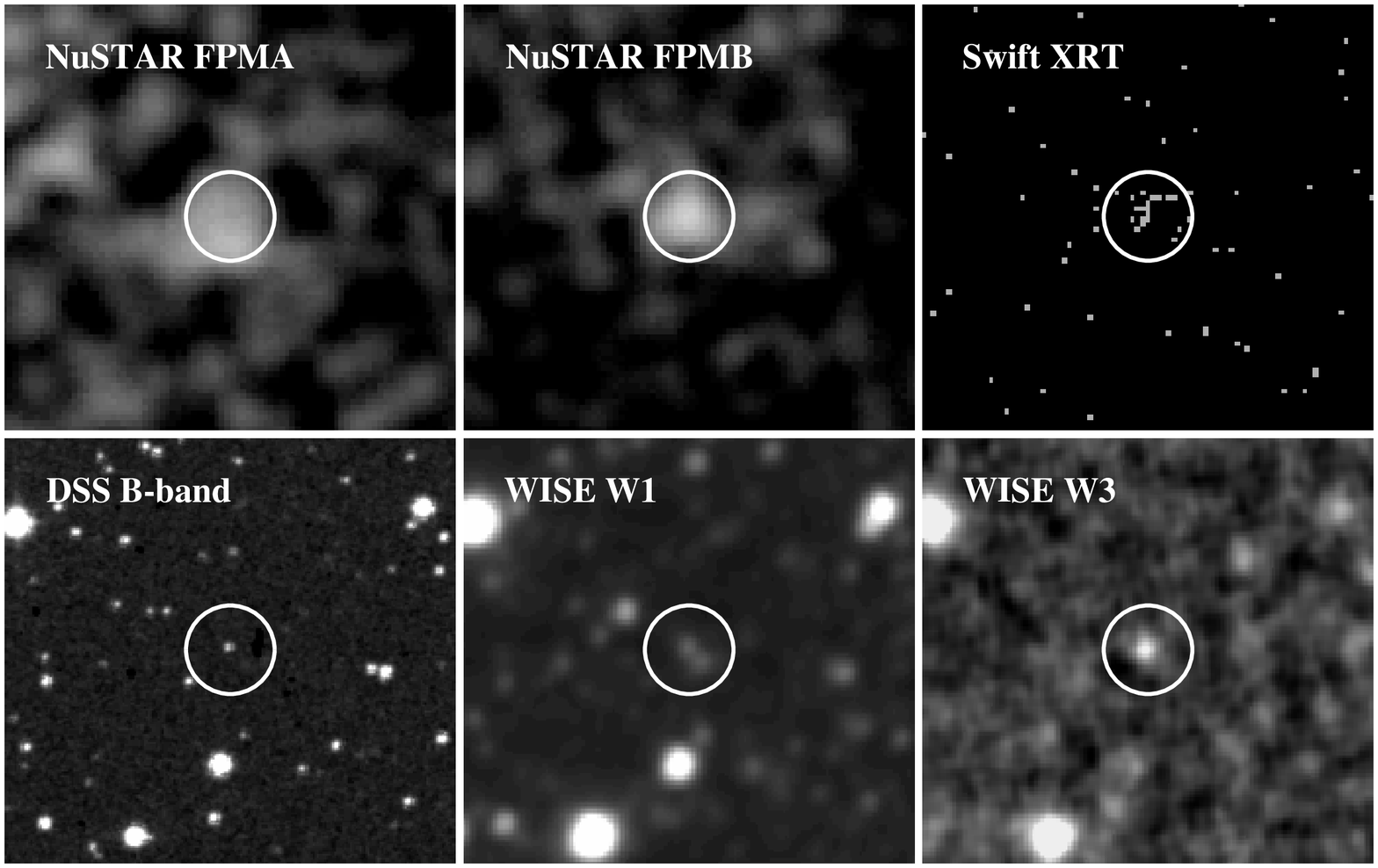}}
\caption{Example multi-wavelength cut-out images to demonstrate some
  of the quality of the multi-wavelength data. The object shown here
  is NuSTAR~J183443+3237.8 in the 3C~382 field, a faint {\it NuSTAR}
  source. The images are (from top left to bottom right): {\it NuSTAR}
  3--24~keV FPMA, {\it NuSTAR} 3--24~keV FPMB, {\it Swift}-XRT
  0.5--10~keV, DSS $B$-band, WISE band 1 (W1; 3.4~$\mu$m), and WISE band 3
  (W3; 12~$\mu$m); the {\it NuSTAR} images have been smoothed with a
  6-pixel ($14{\farcs8}$) Gaussian. The circle has a radius of
  $20^{\prime\prime}$ and is centered on the {\it NuSTAR} source
  position.}
\end{figure*}

\subsubsection{{\it Chandra}, {\it Swift}-XRT, and {\it XMM-Newton} observations}

The archival {\it Chandra} observations are analyzed using CIAO. The
data were reprocessed using the {\tt chandra\_repro} pipeline to
create the new level 2 event file, and the {\it Chandra} source
spectra were extracted from a circular region with a radius of
$\approx$~5$^{\prime\prime}$--10$^{\prime\prime}$. The background
spectra were extracted from several source-free regions of
$\approx$~40$^{\prime\prime}$ radius, selected at different positions
around the source to account for local background variations.

The {\it Swift}-XRT data are reduced using the HEAsoft (v.6.12)
pipeline {\tt xrtpipeline}, which cleans the event files using
appropriate calibration files and extracts the spectra and ancillary
files for a given source position;\footnote{See
  http://heasarc.gsfc.nasa.gov/docs/software/lheasoft/ for details of
  HEAsoft.} the source extraction regions had radii of
$\approx$~20$^{\prime\prime}$. Since the background in the {\it
  Swift}-XRT observations is very low, no background spectra were
extracted.

For the {\it XMM-Newton} EPIC data we used the Pipeline Processing
System (PPS) products, which are a collection of standard processed
high-quality products generated by the Survey Science Center
(SSC). For our analysis we used the \emph{Science Analysis Software}
(SAS v.12.0.1), released in June 2012.\footnote{See
  http://xmm.esa.int/sas/ for details of the SAS software.} After
filtering the event files for high background intervals, we extracted
the source spectra from a circular region with a radius of
$\approx20^{\prime\prime}$. The corresponding background spectra have
been extracted using circular source-free regions in the vicinity of
the corresponding source
($\approx$~30$^{\prime\prime}$--60$^{\prime\prime}$ radius
regions). Using the SAS tasks {\tt rmfgen} and {\tt arfgen} we also
produced the response matrices for each source in each of the three
EPIC cameras separately (pn, MOS1, and MOS2).

%
%
\begin{figure*}
\includegraphics[angle=0,width=18.0cm]{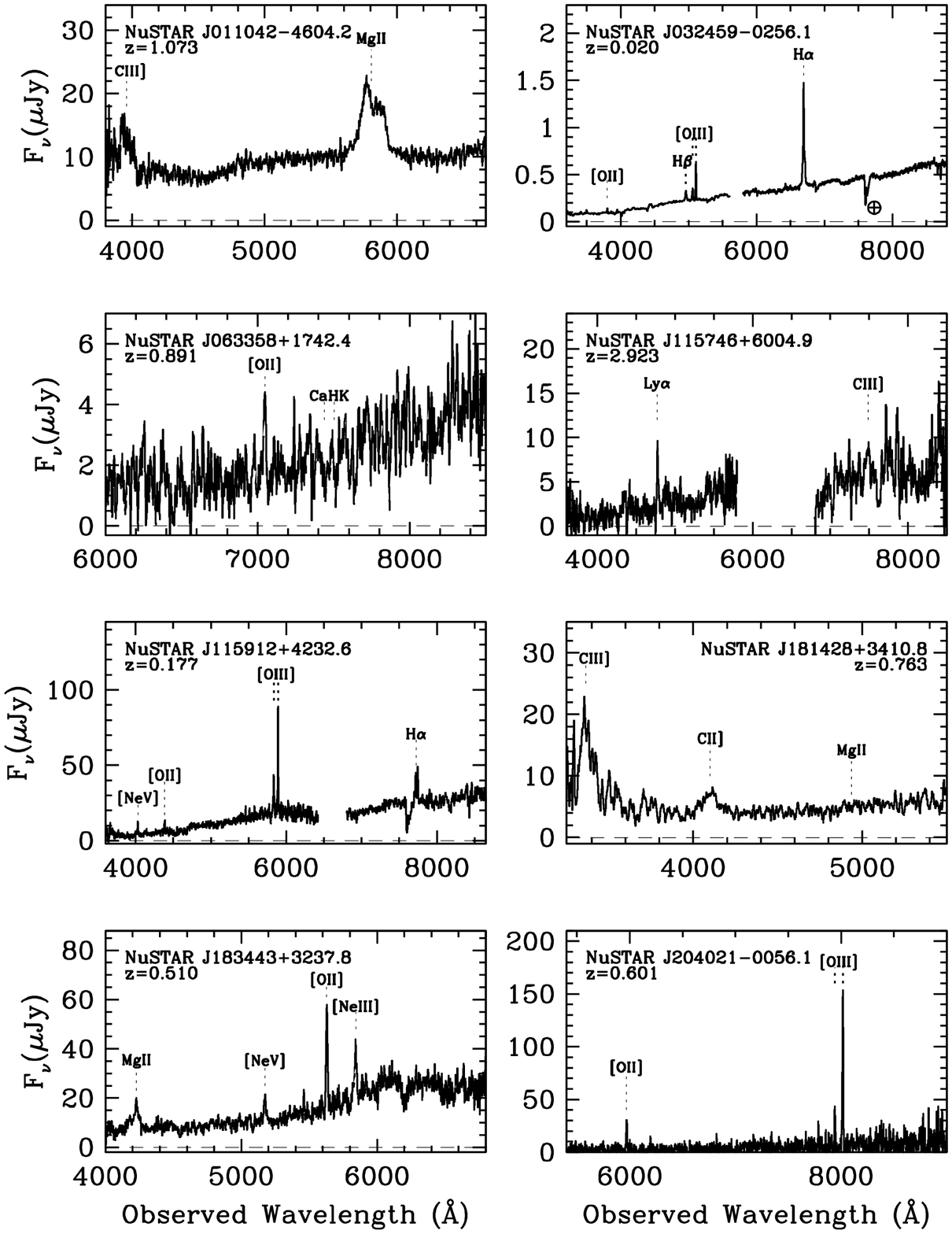}
\caption{Optical spectra for the eight newly identified serendipitous
  {\it NuSTAR} sources; the optical spectra of the other two sources
  (NuSTAR~J121027+3929.1 and NuSTAR~J145856-3135.5) have been
  previously presented in Morris et~al. (1991) and Caccianiga
  et~al. (2008). The prominent emission and absorption lines are
  indicated; see \S2.5.}
\end{figure*}

\subsection{Counterpart matching}

To provide reliable source identification we matched the {\it NuSTAR}
sources to the $<10$~keV and multi-wavelength data; see \S2.2, \S2.4,
and Table~3 for the description of the data. We searched for
multi-wavelength counterparts within $10^{\prime\prime}$ of the {\it
  NuSTAR} source positions using on-line source catalogs and
multi-wavelength images; the latter approach is required for faint
counterparts or for recent data not yet reported in on-line source
catalogs. The $10^{\prime\prime}$ search radius is motivated by the
absolute astrometric accuracy of {\it NuSTAR} ($\pm5^{\prime\prime}$,
90\% confidence, for bright X-ray sources; Harrison et~al. 2013) and
the low count rates for the majority of our sources.

A lower-energy X-ray counterpart is found within $10^{\prime\prime}$
for each of the {\it NuSTAR} sources; see Table~2. To provide further
confidence that the X-ray source is the correct lower-energy
counterpart to the {\it NuSTAR} source, we compared the 3--8~keV
fluxes of the lower-energy source and the {\it NuSTAR} source. We
selected and extracted the lower-energy X-ray data following \S2.2 and
we calculated the 3--8~keV fluxes using a power-law model in {\sc
  xspec} (the model component is {\sc pow} in {\sc xspec}); see
Table~1 for details of the low-energy X-ray data selected for each
source. The average source flux was calculated for the {\it
  XMM-Newton} data when multiple detectors were used (i.e.,\ PN, MOS~1
and MOS~2). In Fig.~1 we compare the 3--8~keV fluxes from the
lower-energy X-ray data to the 3--8~keV flux from the {\it NuSTAR}
data. In all cases the fluxes agree within a factor of two,
demonstrating that we have selected the correct lower-energy X-ray
counterpart.

An optical counterpart is also found within $10^{\prime\prime}$ of
each {\it NuSTAR} source; see Table~3. Given the larger intrinsic
uncertainty in the {\it NuSTAR} source position when compared to the
lower-energy X-ray source position, we also measured the distance
between the lower-energy X-ray source position and the optical
position. An optical counterpart is found within $3^{\prime\prime}$
(and the majority lie within $1^{\prime\prime}$) of the lower-energy
X-ray source position for all of the sources. See Fig.~2 for example
multi-wavelength cut-out images of NuSTAR~J183443+3237.8 in the 3C~382
field.

\subsection{Ultraviolet--radio data}

To further characterize the properties of the {\it NuSTAR} sources we
used ultraviolet (UV) to mid-infrared (MIR) data. Table~3 presents the
broad-band UV--MIR photometric properties of the {\it NuSTAR} sources,
primarily obtained from existing, publicly available all-sky or
large-area surveys, including the {\it Galaxy Evolution Explorer}
({\it GALEX}; Martin et~al. 2005), the Digitised Sky Survey (DSS;
Minkowski \& Abell 1963; Hambly et~al. 2001), the Sloan Digital Sky
Survey (SDSS; York et~al. 2000), the Two-Micron All-Sky Survey (2MASS;
Skrutskie et~al. 2006), and the {\it Wide-Field Infrared Survey
  Explorer} ({\it WISE}; Wright et~al. 2010). The source photometry is
provided in its native format for all of the sources. The DSS data,
provided for sources outside of the SDSS, were obtained from the
SuperCOSMOS scans of the photographic Schmidt plates (Hambly
et~al. 2001). As recommended by the SuperCOSMOS Sky Survey, all
photometric uncertainties are set to 0.30~mag for those measurements.
Where publicly available, we also provide {\it Spitzer} photometry
from the Infrared Array Camera (IRAC; Fazio et~al. 2004), obtained
from the post-basic calibrated data (PBCD) products.  To avoid the
effects of source confusion, photometry was measured in 2\farcs4
radius apertures on the 0\farcs6 per pixel re-sampled PBCD mosaics,
and then corrected to total flux density using aperture corrections
from the IRAC Instrument Handbook (v.2.0.2).\footnote{See
  http://irsa.ipac.caltech.edu/data/SPITZER/docs/irac/ for details of
  {\it Spitzer}-IRAC.} Several sources were observed during the
post-cryogenic {\it Warm Spitzer} phase, and thus only the two shorter
wavelength bandpasses from {\it Spitzer}-IRAC are available.

In several cases we used photometry from different sources, which we
list below. For NuSTAR~J063358+1742.4 we report a $J$-band
non-detection, which is measured from 1.56~ks of dithered observations
obtained with the Florida Infrared Imaging Multi-object Spectrograph
(FLAMINGOS) on the Kitt Peak 2.1~m telescope. The data were obtained
on UT 2012 October 17 in photometric but 1\farcs6 seeing conditions,
and the 3~$\sigma$ upper limit was calculated in a $2^{\prime\prime}$
radius aperture; see Table~3 for more details. For
NuSTAR~J145856-3135.5 we report the $R$-band magnitude from Caccianiga
et~al. (2008). For NuSTAR~J181428+3410.8 the optical photometry comes
from imaging reported in Eisenhardt et~al. (2012), calibrated to the
SDSS. The {\it WISE} 12~$\mu$m photometry for NuSTAR~J181428+3410.8
was measured directly from the images as this source does not appear
in the {\it WISE} All-Sky Catalog; we do not provide the shorter
wavelength {\it WISE} photometry for this source as it is superceded
by {\it Warm Spitzer} observations. For NuSTAR~J183443+3237.8 we
obtained $B$, $R$, and $I$ band observations using the Palomar 60-inch
telescope (P60) on UT 2013 March 04 in $\approx2^{\prime\prime}$
seeing; the exposure time was 300~s in each band, repeated three times
with a $60^{\prime\prime}$ dither. NuSTAR~J183443+3237.8 was well
detected in all three bands and the reported photometry in Table~3 was
measured in $4^{\prime\prime}$ diameter apertures, which has been
corrected for PSF losses.

We also searched for radio counterparts in the NVSS and FIRST VLA
surveys (Becker et~al. 1995; Condon et~al. 1998), using a search
radius of $30^{\prime\prime}$ and $15^{\prime\prime}$,
respectively. NuSTAR~J121027+3929.1 was detected in both surveys and
has a flux of $f_{\rm 1.4 GHz}=18.7\pm0.7$~mJy (in the NVSS survey),
which corresponds to a rest-frame luminosity density of $L_{\rm 1.4
  GHz}=2.2\times10^{24}$~W~Hz$^{-1}$ (calculated following Equation 2
of Alexander et~al. 2003b and assuming a radio spectral slope of
$\alpha=0.8$). With the exception of NuSTAR~J011042-4604.2, all of the
other sources had at least NVSS coverage but none were detected. The
rest-frame luminosity density upper limits ranged from $L_{\rm 1.4
  GHz}<1.8\times10^{20}$~W~Hz$^{-1}$ (for NuSTAR~J032459-0256.1) to
$L_{\rm 1.4 GHz}<4.3\times10^{24}$~W~Hz$^{-1}$ (for
NuSTAR~J115746+6004.9), with the majority of the sources having upper
limits of $L_{\rm 1.4 GHz}<10^{23}$--$10^{24}$~W~Hz$^{-1}$.

\subsection{Optical spectroscopy}

Two of the ten serendipitous sources have existing optical
spectroscopy: NuSTAR~J121027+3929.1 has been previously identified as
a BL Lac at $z=0.615$ (MS~1207.9+3945; e.g.,\ Stocke et~al. 1985;
Gioia et~al. 1990; Morris et~al. 1991) while NuSTAR~J145856-3135.5 has
been previously identified as a broad-line AGN (BLAGN) at $z=1.045$
(2XMM~J145857.0-313536; Caccianiga et~al. 2008). For the other eight
serendipitous {\it NuSTAR} sources we obtained optical spectroscopy at
the Palomar, Keck, and Gemini-South telescopes. Table~3 presents basic
information about the observations, including the instrument and UT
date of the observations and in the Appendix we provide specific
details for each observation. We processed all of the optical
spectroscopic data using standard techniques, and flux calibrated the
spectra using standard stars observed on the same nights.

The optical spectra for the eight newly identified {\it NuSTAR}
sources are shown in Fig.~3. Clear multiple broad and/or narrow
emission lines are detected in six sources, showing that the redshift
identifications are reliable. However, the optical counterparts for
NuSTAR~J115746+6004.9 and NuSTAR~J063358+1742.4 are comparatively
faint and the optical spectra are therefore of lower quality when
compared to the optical spectra of the other serendipitous
sources. NuSTAR~J115746+6004.9 has narrow, spatially extended
Ly~$\alpha$ emission as well as somewhat broadened \ion{C}{3}]
  emission indicating $z=2.923$; spatially extended Ly~$\alpha$
  emission is often found to be associated with powerful AGNs
  (e.g.,\ Reuland et~al. 2003; Geach et~al. 2009; Yang
  et~al. 2009). The redshift of NuSTAR~J063358+1742.4 is less certain
  due to the identification of a single narrow emission line, which is
  more likely to be [\ion{O}{2}] at $z=0.891$ than Ly~$\alpha$ due to
  the rising optical continuum and lack of a strong Ly~$\alpha$ forest
  decrement (as would be expected had the source been at $z \sim
  4.8$); the identification of two absorption features at the
  wavelengths expected for Ca~H+K provide additional confidence for
  $z=0.891$. We consider all of the redshifts to be reliable.

The two {\it NuSTAR} sources with existing optical spectroscopy
(NuSTAR~J121027+3929.1; NuSTAR~J145856-3135.5) have optical magnitudes
consistent with the eight newly identified {\it NuSTAR} sources and
meet our basic requirement for inclusion in this paper (i.e.,\ sources
identified in {\it NuSTAR} observations taken up until January 31st
2013); we note that several of the other $\approx$~40 serendipitously
detected {\it NuSTAR} sources also have existing optical spectroscopy
but have been identified in more recent {\it NuSTAR} observations and
so are not included in this paper. We therefore believe that the
inclusion of these two {\it NuSTAR} sources does not bias our overall
{\it NuSTAR} sample.

%
\section{Data Analysis}
%

%
%
\begin{figure*}[!t]
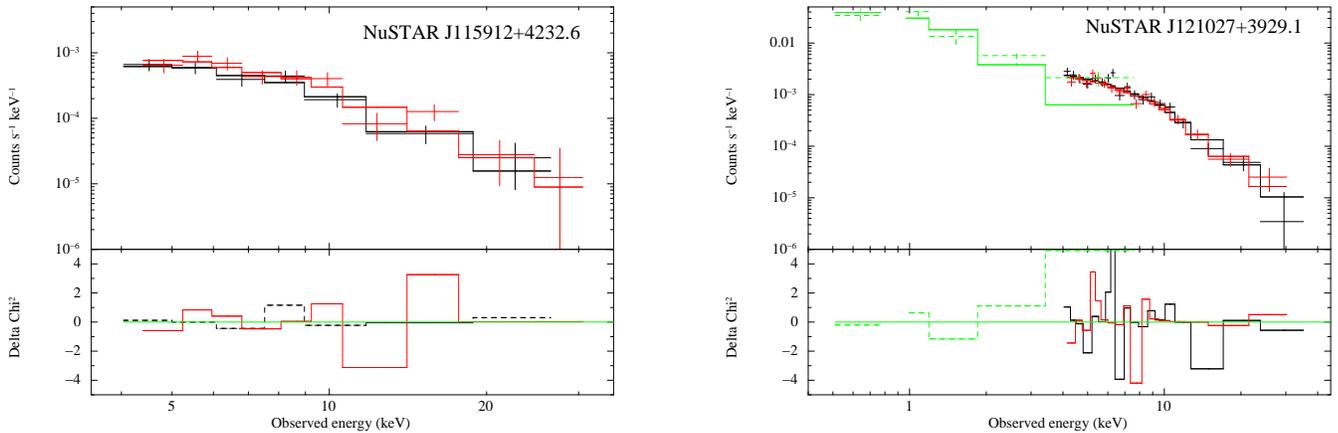

{\includegraphics[angle=-90,width=8.5cm]{dalexander_fig4a.eps}
\hfill
\includegraphics[angle=-90,width=8.5cm]{dalexander_fig4b.eps}}
\caption{Example X-ray spectra and best-fitting power-law model
  solutions for NuSTAR~J115912+4232.6 (left) and NuSTAR~J121027+3929.1
  (right). The {\it NuSTAR} data is plotted for NuSTAR~J115912+4232.6
  over 4--32~keV and the {\it NuSTAR} and {\it Swift}-XRT data is
  plotted for NuSTAR~J121027+3929.1 over 0.5--50~keV. The black and
  red crosses are from {\it NuSTAR} FPMA and FPMB, respectively, and
  the green crosses are from {\it Swift}-XRT. The X-ray data have been
  grouped and fitted using a power-law model and $\chi^2$ statistics;
  see \S3.1. The best-fitting models are plotted as solid lines and
  the bottom panels show the deviations of the data from the
  best-fitting model ($\Delta$~$\chi^2$).}
\end{figure*}

\subsection{X-ray spectral fitting}

To interpret the X-ray data and provide insight into the intrinsic AGN
properties of the serendipitous {\it NuSTAR} sources (e.g.,\ $\Gamma$
and $N_{\rm H}$) we fitted the X-ray data using physically motivated
AGN models. We extracted the {\it NuSTAR} data following \S2.1.5 and
the lower-energy X-ray data following \S2.2.1.

For the three sources with $>200$~counts in each {\it NuSTAR} FPM at
3--24~keV (NuSTAR~J011042-4604.2, NuSTAR~J115912+4232.6, and
NuSTAR~J121027+3929.1; see Table~2), we grouped the {\it NuSTAR} data
into bins of at least 40 counts per bin and used $\chi^2$ statistics
to find the best-fitting model parameter solutions. However, the {\it
  NuSTAR} photon statistics were too poor to allow for $\chi^2$
statistics for the other seven sources, and for the X-ray spectral
analyzes of these sources we fitted the unbinned X-ray data using the
$C$-statistic (Cash 1979). The $C$-statistic is calculated on unbinned
data and is therefore ideally suited to low-count sources
(e.g.,\ Nousek \& Shue 1989). However, since the data need to be
fitted without the background subtracted, it is essential to
accurately characterize the background and use that as a fixed model
component in the X-ray spectral fitting of the source spectrum. We
characterized the background by fitting the background regions using a
double power-law model (the model components are {\sc pow}*{\sc pow}
in {\sc xspec}). The photon statistics were also often poor for the
lower-energy X-ray data ($<200$~counts) and we therefore typically
fitted the unbinned $<10$~keV data using the $C$-statistic with the
measured background as a fixed component. In Fig.~4 we show example
{\it NuSTAR} spectra for two of the brightest {\it NuSTAR} sources:
NuSTAR~J115912+4232.6 and NuSTAR~J121027+3929.1; for
NuSTAR~J121027+3929.1 we also show the {\it Swift}-XRT data. All fit
parameter uncertainties are quoted at the 90\% confidence level (Avni
1976).

We initially fitted only the {\it NuSTAR} data using a simple
power-law model (the {\sc pow} model in {\sc xspec}) to provide
constraints on the overall X-ray spectral slope ($\Gamma$) over
4--32~keV. We also restricted the {\it NuSTAR} data to cover the
rest-frame 10--40~keV energy range for each source and fitted a
power-law model to measure both the rest-frame 10--40~keV spectral
slope ($\Gamma_{\rm 10-40 keV}$) and luminosity ($L_{\rm 10-40 keV}$);
given the redshift of NuSTAR~J115746+6004.9 ($z=2.923$) we fitted to
the rest-frame 15--60~keV data. See Table~4.

To provide direct measurements on the presence of absorption we
jointly fitted an absorbed power-law model (the model components are
{\sc zwabs*pow} in {\sc xspec}) to both the {\it NuSTAR} and
lower-energy X-ray data for each source.\footnote{We note that AGNs
  often require more complex models to characterize their X-ray
  emission than that of a simple absorbed power law (e.g.,\ Winter
  et~al. 2009; Vasudevan et~al. 2013). However, the data quality of
  our sources is not sufficient to reliably constrain such models on a
  source by source basis (see \S4.3 for more detailed average
  constraints).} For five of the sources we fitted the 0.5--32~keV
data (for NuSTAR~J121027+3929.1 we fitted the 0.5--50~keV data, given
the good photon statistics of this source), jointly fitting the X-ray
spectral slope and absorbing column density for both of the {\it
  NuSTAR} FPMs and the lower-energy X-ray data. However, for
NuSTAR~J115912+4232.6 no good-quality low-energy X-ray data exist and
we therefore only fitted the {\it NuSTAR} data, while for the
remaining three sources (NuSTAR~J115746+6004.9, NuSTAR~J145856-3135.5,
and NuSTAR~J181428+3410.8) the photon statistics of the {\it NuSTAR}
data were too poor to provide reliable constraints on both $\Gamma$
and $N_{\rm H}$, and we therefore fitted the absorbed power-law model
to just the lower-energy X-ray data. The best-fitting model parameters
are given in Table~4.

%
%
\begin{figure*}
\centerline{\includegraphics[angle=0,width=18.0cm]{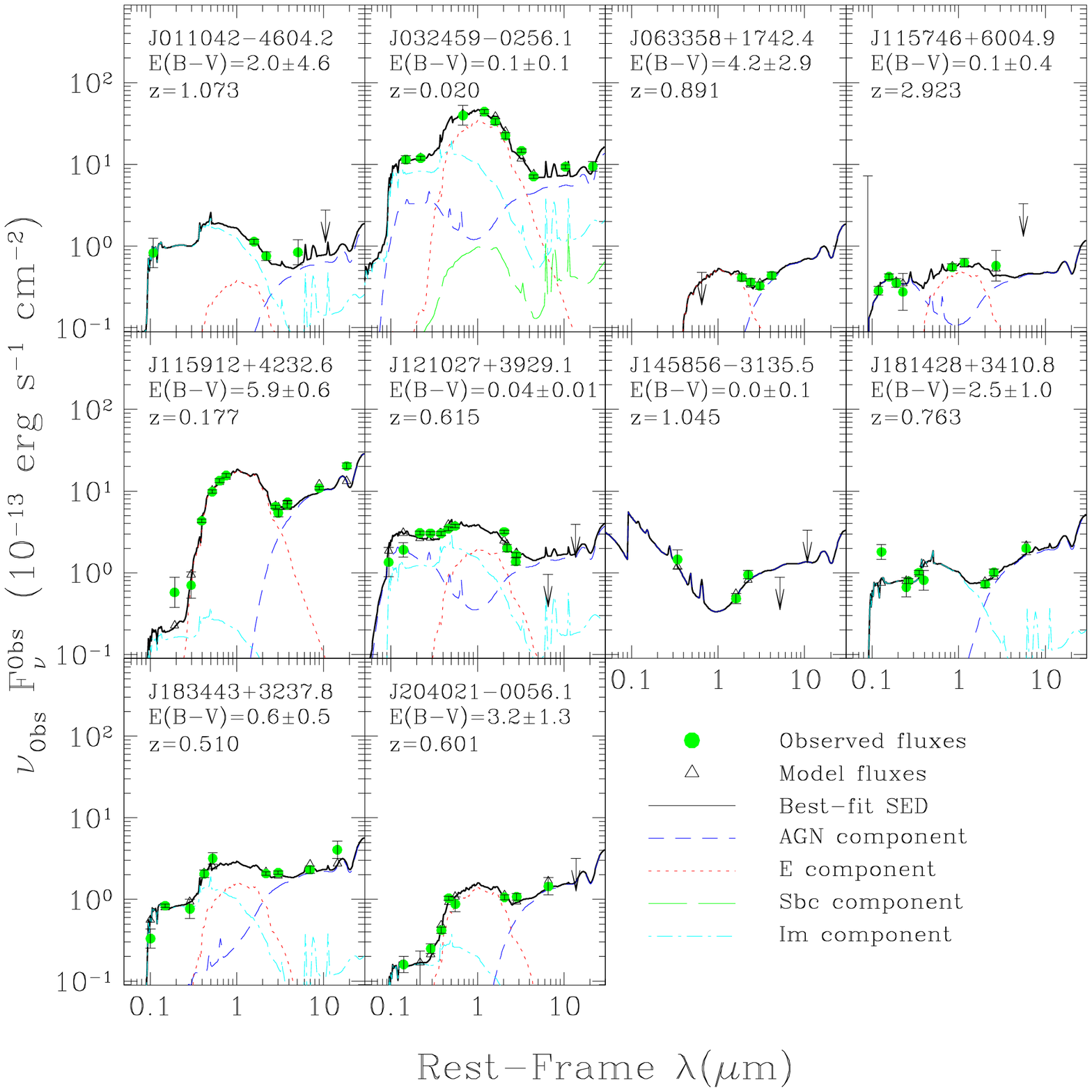}}
\caption{Broad-band UV--MIR SED and best-fitting model solution for
  the serendipitous {\it NuSTAR} sources. The data are fitted with the
  Assef et~al. (2010) AGN (magenta dashed curve) and galaxy
  (elliptical: red dotted curve; spiral: green long-dashed curve;
  irregular: cyan dash-dotted curve) templates. The best-fitting
  solution is plotted as a black solid curve. The source redshift,
  best-fitting dust-reddening solution ($E(B-V)$) and uncertainties
  are shown.}
\end{figure*}

\subsection{Ultraviolet--mid-infrared spectral energy distribution fitting}

To constrain the relative contributions from AGN activity and the host
galaxy to the UV--MIR data we fitted the broad-band UV--MIR spectral
energy distributions (SEDs) using the 0.03--30~$\mu$m empirical
low-resolution templates for AGN and galaxies of Assef
et~al. (2010). Each SED is modelled as the best-fit non-negative
combination of three galaxy templates and an AGN template. The
reddening of the AGN template, parameterized by $E(B-V)$, is a free
parameter in the fit. The errors on the parameters were calculated
using a Monte-Carlo method, where the photometry is resampled 1000
times according to the photometric uncertainties and the SED fits and
parameters are re-calculated; the errors refer to the standard
deviation for all of the realizations. Since the templates have been
empirically defined using AGNs with similar X-ray luminosities and
redshifts at the {\it NuSTAR} sources, we do not expect there to be
significant systematic uncertainties in the best-fitting model
solutions; the efficacy of the SED-fitting approach will be further
explored in S.~M.~Chung et~al. (in prep.). We refer the reader to
Assef et~al. (2008, 2010, 2013) for further details.

In Fig.~5 we present the UV--MIR SEDs and best-fitting solutions and
in Table~3 we provide the following best-fitting parameters: $\hat{a}$
(the fractional contribution to the overall emission from the AGN component over
0.1--30~$\mu$m; Assef et~al. 2013), $E(B-V)$ (the dust reddening of the AGN component),
$L_{\rm 6\mu m}$ (the luminosity of the AGN component at rest-frame
6~$\mu$m), and $M_{*}$ (the stellar mass of the host galaxy). The
stellar mass is calculated from the absolute magnitude of the stellar
component using the color-magnitude calibration of Bell
et~al. (2003). Three of the {\it NuSTAR} sources have photometric
measurements in $\le$~5 bands (NuSTAR~J11042+4604.2;
NuSTAR~J063358+1742.4; NuSTAR~J145856-3135.5) and the derived
properties for these sources are therefore poorly constrained.

%
\section{Results}
%

In analyzing the {\it NuSTAR} sources we predominantly focus on
characterizing their X-ray and UV--MIR properties and comparing these
properties to those of sources detected in previous-generation
$\simgt10$~keV surveys (e.g.,\ {\it Swift}-BAT; Tueller et~al. 2008,
2010; Baumgartner et~al. 2012).

%
%
\begin{figure}[!t]
\centerline{\includegraphics[angle=0,width=8.0cm]{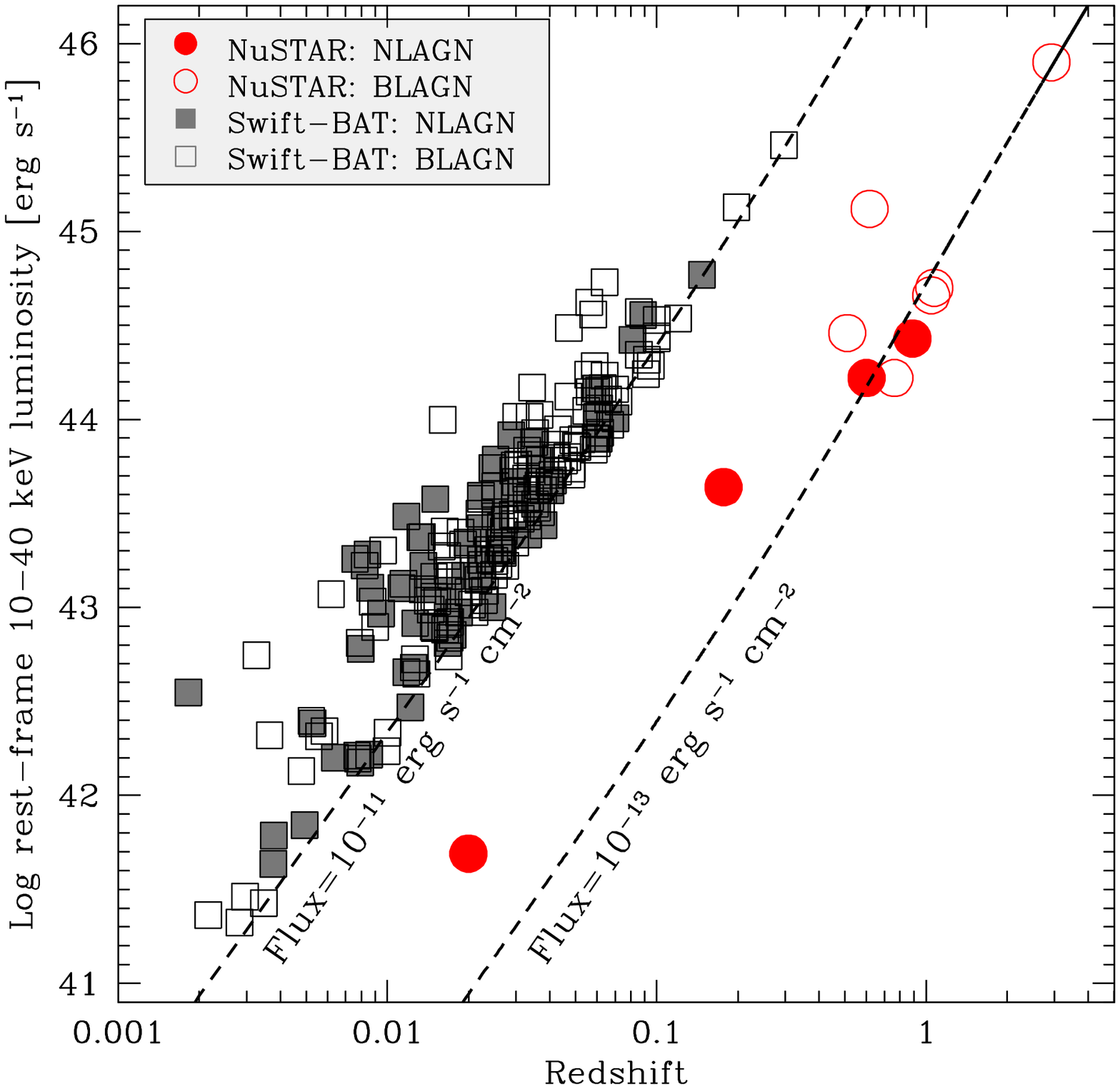}}
\caption{Rest-frame 10--40~keV luminosity versus redshift for the {\it
    NuSTAR} sources (circles) compared to the {\it Swift}-BAT AGN
  sample of Burlon et~al. (2011; squares); filled symbols indicate
  narrow-line AGNs (NLAGN) and open symbols indicate broad-line AGNs
  (BLAGN). The rest-frame 10--40~keV luminosity for the {\it NuSTAR}
  sources is calculated directly from the X-ray spectra (see Table~4)
  while the rest-frame 10--40~keV luminosity for the {\it Swift}-BAT
  AGNs is calculated from the observed-frame 15--55~keV flux, assuming
  $\Gamma=1.8$ for the $K$-correction factor. The dashed lines
  indicate different flux limits and show that the {\it NuSTAR}
  sources are up-to $\approx$~100 times fainter than the {\it
    Swift}-BAT AGNs.}
\end{figure}

\subsection{Basic source properties}

The 8--24~keV fluxes of the {\it NuSTAR} sources are up-to
$\approx$~100 times fainter than sources previously detected at
$\simgt$~10~keV ($f_{\rm 8-24
  keV}\approx$~(0.6--5.9)~$\times10^{-13}$~erg~s$^{-1}$~cm$^{-2}$, as
compared to $f_{\rm 8-20
  keV}\simgt0.4\times10^{-11}$~erg~s$^{-1}$~cm$^{-2}$; e.g.,\ see
Table~2 and the {\it RXTE} data in Revnivtsev et~al. 2004). The {\it
  NuSTAR} sources also have fainter optical counterparts and lie at
higher redshifts than sources previously detected at $\simgt$~10~keV
($R\approx$~16--22~mags and a median redshift of $z\approx$~0.7, as
compared to $V\approx$~10--16~mags and a median redshift of
$z\approx$~0.03; see Beckmann et~al. 2009 and Table~3).

%
%
\begin{figure}[!t]
\centerline{\includegraphics[angle=0,width=8.0cm]{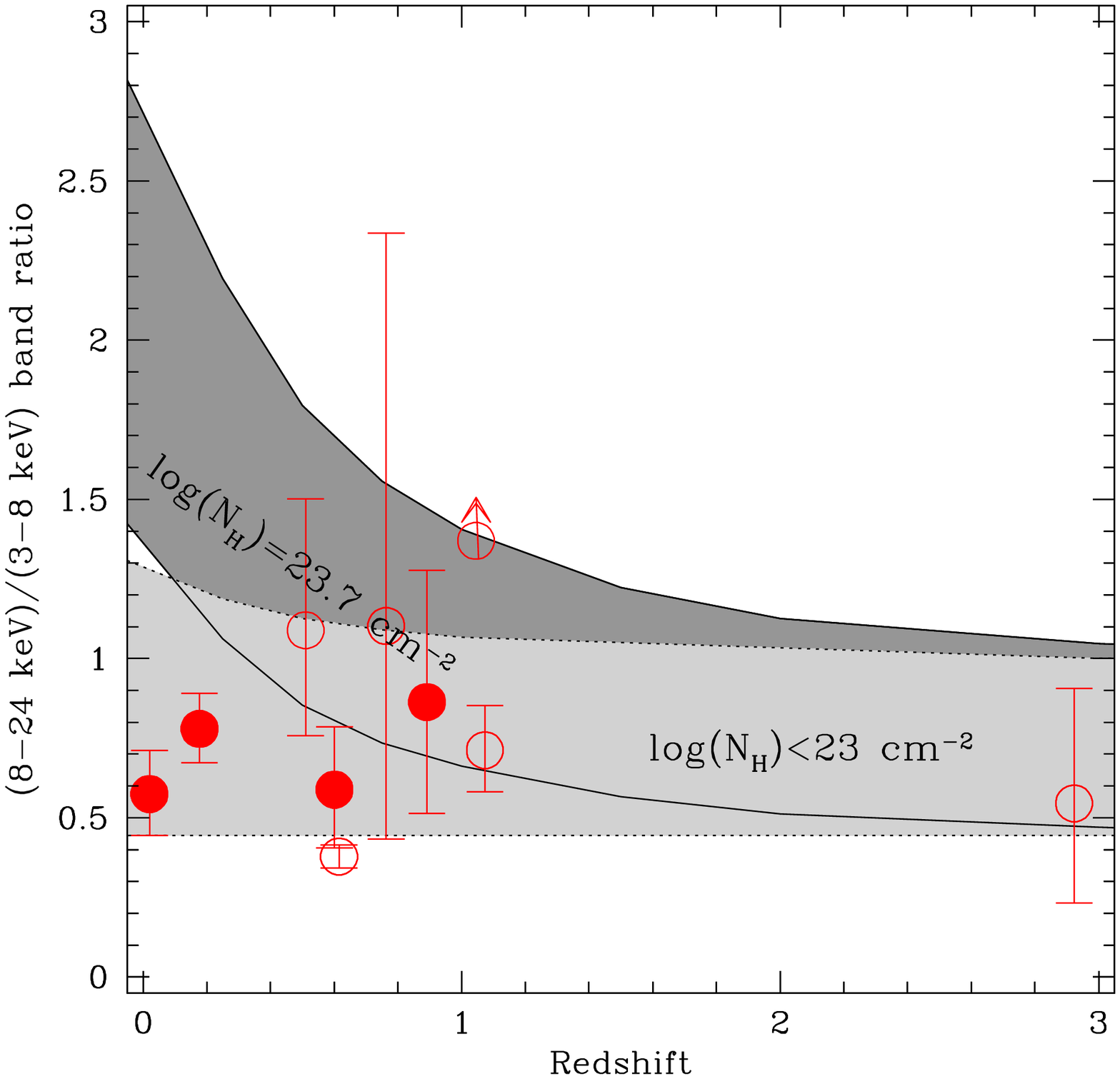}}
\caption{X-ray band ratio versus redshift for the {\it NuSTAR}
  sources; see Fig.~6 for a description of the symbols. The error bars
  indicate the 1~$\sigma$ uncertainty on the band ratio, which is
  calculated following the ``numerical method'' in \S1.7.3 of Lyons
  (1991). The shaded regions show the range of expected band ratios
  for AGNs with $N_{\rm H}<10^{23}$~cm$^{-2}$ and $N_{\rm
    H}\approx5\times10^{23}$~cm$^{-2}$, for an intrinsic spectral
  slope of $\Gamma=1.8\pm0.5$; the dotted and solid curves indicate
  the maximum extents in band ratio for $N_{\rm H}<10^{23}$~cm$^{-2}$
  and $N_{\rm H}=5\times10^{23}$~cm$^{-2}$, respectively. The
  predicted band ratios were calculated using the RMF and ARF for
  NuSTAR~J183443+3237.8 (see \S2.1.5).}
\end{figure}

In Fig.~6 we plot the rest-frame 10--40~keV luminosity versus redshift
of the {\it NuSTAR} sources and compare them to AGNs detected in the
{\it Swift}-BAT survey (e.g.,\ Burlon et~al. 2011). With a median
luminosity of $L_{\rm 10-40 keV}\approx3\times10^{44}$~erg~s$^{-1}$,
the {\it NuSTAR} sources are more luminous than the vast majority of
the {\it Swift}-BAT AGNs, where $\approx$~80\% have $L_{\rm 10-40
  keV}<10^{44}$~erg~s$^{-1}$; the median luminosity of the {\it
  Swift}-BAT AGNs is $L_{\rm 10-40
  keV}\approx3\times10^{43}$~erg~s$^{-1}$. The larger fraction of
luminous AGNs detected by {\it NuSTAR}, in comparison to {\it
  Swift}-BAT, is a consequence of the higher sensitivity of {\it
  NuSTAR} and two additional factors (1) the strong redshift-dependent
evolution of luminous AGNs (e.g.,\ Ueda et~al. 2003; Barger
et~al. 2005; Hasinger et~al. 2005; Aird et~al. 2010), and (2) the
comparatively small cosmological volume in which {\it NuSTAR} is
sensitive to AGNs with $L_{\rm 10-40 keV}<10^{44}$~erg~s$^{-1}$
($z\simlt0.2$).

The range of redshifts for the {\it NuSTAR} sources is large
($z=$~0.020--2.923). At $z=2.923$, NuSTAR~J115746+6004.9 is the
highest-redshift AGN detected to date at $\simgt10$~keV that does not
appear to be strongly beamed (e.g.,\ Beckmann et~al. 2009; Burlon
et~al. 2011; Malizia et~al. 2012). By comparison,
NuSTAR~J032459-0256.1 has a redshift typical of those of the {\it
  Swift}-BAT AGNs ($z=0.020$) but, with $L_{\rm 10-40
  keV}\approx5\times10^{41}$~erg~s$^{-1}$, it is $\approx$~30 times
less luminous than the faintest {\it Swift-BAT} AGNs; in \S4.4 we show
that this source is also unusual since it is hosted in a low-mass
dwarf galaxy. The high X-ray luminosities for the majority of the {\it
  NuSTAR} sources indicate that they are AGNs. However, the origin of
the modest X-ray luminosity of NuSTAR~J032459-0256.1 is less clear and
it is possible that the X-ray emission is produced by a hyper-luminous
X-ray source (HLX; e.g.,\ Farrell et~al. 2009; Swartz et~al. 2011) as
opposed to a low-luminosity AGN; high-spatial resolution observations
with {\it Chandra} would be able to distinguish between an off-nuclear
HLX and an AGN or nuclear HLX. The median and range in X-ray
luminosity and redshift of the {\it NuSTAR} sources are consistent
with expectations (Ballantyne et~al. 2011). However, we note that both
the redshift and X-ray luminosity of NuSTAR~J032459-0256.1 are below
the range typically explored in the models.

The optical spectral properties of the {\it NuSTAR} sources are
relatively diverse; see Fig.~3 and Table~3. Five of the ten
($\approx50^{+34}_{-22}$\%) serendipitous sources have broad emission
lines and are classified as broad-line AGNs (BLAGNs), four
($\approx40^{+32}_{-19}$\%) have narrow emission lines and we classify
as narrow-line AGNs (NLAGNs), and one is a BL Lac, with strong
power-law optical continuum emission and weak emission
lines.\footnote{All errors are taken from Tables 1 and 2 of Gehrels
  (1986) and correspond to the 1~$\sigma$ level; these were calculated
  assuming Poisson statistics.}$^,$\footnote{We note that our
  classification of NLAGNs is fairly loose since we lack the
  emission-line diagnostics around $H\alpha$ for the majority of our
  sources to prove that they lie in the AGN region of an emission-line
  diagnostic diagram as opposed to the HII region (e.g.,\ Baldwin,
  Phillips, \& Terlevich 1981; Veilleux \& Osterbrock 1987).} The BL
Lac (NuSTAR~J121027+3929.1) is a relatively well studied
high-frequency peaked BL Lac (HBL; Padovani \& Giommi 1995),
originally identified at X-ray energies by {\it Einstein}
(MS~1207.9+3945; e.g.,\ Gioia et~al. 1990; Morris et~al. 1991; Urry
et~al. 2000; Maselli et~al. 2008). Two of the NLAGNs have $L_{\rm
  10-40 keV}>10^{44}$~erg~s$^{-1}$ and are therefore type 2 quasars,
representing $\approx20^{+26}_{-13}$\% of the {\it NuSTAR} sample; by
comparison six type 2 quasars are identified in the 199 {\it
  Swift}-BAT sample of Burlon et~al. (2011), just
$\approx3^{+2}_{-2}$\% of the entire sample. However, the difference
in the fraction of type 2 quasars between {\it NuSTAR} and {\it
  Swift}-BAT is at least partly related to the increased fraction of
luminous AGNs in the {\it NuSTAR} serendipitous sample; we note that,
since we lack coverage of the $H\alpha$ emission line for the type 2
quasars, we cannot rule out the presence of broad $H\alpha$ in some of
the {\it NuSTAR} type 2 quasars. The overall fraction of BLAGNs and
NLAGNs in the {\it Swift}-BAT AGN sample is consistent with that found
for the {\it NuSTAR} serendipitous sample: $\approx50^{+5}_{-5}$\% of
the {\it Swift}-BAT sources are BLAGNs (including all Seyfert 1s and
Seyfert 1.2s) and $\approx50^{+5}_{-5}$\% are NLAGNs (including all
Seyfert 1.5s, Seyfert 1.8s, Seyfert 1.9s, and Seyfert 2s). Therefore,
within the limitations of our small sample, the biggest differences
between the basic properties of the {\it NuSTAR} sources and the {\it
  Swift}-BAT AGNs appear to be luminosity and redshift.

\subsection{X-ray spectral properties: the presence of absorption}

The $\simgt10$~keV sensitivity of {\it NuSTAR} allows for the
selection of AGNs almost irrespective of the presence of absorption,
up-to high absorbing column densities of $N_{\rm
  H}\approx$~(1--3)~$\times10^{24}$~cm$^{-2}$. However, particularly
when using lower-energy X-ray data, we can measure the absorbing
column densities of the {\it NuSTAR} sources using the X-ray band
ratio (the 8--24~keV to 3--8~keV count-rate ratio) and from fitting
the X-ray spectra over a broad energy range.

In Fig.~7 we show the X-ray band ratio versus redshift for the {\it
  NuSTAR} sources and compare them with those expected for absorbed
power-law emission from an AGN. As can be seen, given the high X-ray
energies probed by {\it NuSTAR}, the evidence for absorption can only
be clearly identified on the basis of the X-ray band ratio for the
most heavily obscured AGNs ($N_{\rm H}\simgt5\times10^{23}$~cm$^{-2}$)
at $z\simlt$~0.5. The X-ray band ratios for all of the {\it NuSTAR}
sources are consistent with $N_{\rm
  H}\simlt5\times10^{23}$~cm$^{-2}$. However, more detailed
constraints on the X-ray spectral properties and the presence of
absorption can be placed by directly fitting the X-ray spectra of the
{\it NuSTAR} sources, particularly when including lower-energy data
($\simlt3$~keV), which is more sensitive to column densities of
$N_{\rm H}\simlt10^{23}$~cm$^{-2}$. We extracted the X-ray spectral
products and fitted the X-ray data of the {\it NuSTAR} sources with an
absorbed power-law model ({\sc zwabs*pow} in {\sc xspec}), following
\S3.1; see Footnote 7 for caveats on the application of an absorbed
power-law model to characterize AGNs. In Fig.~8 we plot the
best-fitting X-ray spectral slope ($\Gamma$) and absorbing column
density ($N_{\rm H}$) for the {\it NuSTAR} sources (see Table~4 for
the best-fitting parameters) and compare them to the X-ray spectral
properties of the {\it Swift}-BAT-detected AGNs in Burlon
et~al. (2011). The best-fitting X-ray spectral slopes of the {\it
  NuSTAR} sources are broadly consistent with those found for
well-studied nearby AGNs ($\Gamma\approx$~1.3--2.3; e.g.,\ Nandra \&
Pounds 1994; Reeves \& Turner 2000; Deluit \& Courvoisier 2003;
Piconcelli et~al. 2005; Burlon et~al. 2011). The source with the
steepest X-ray spectral slope ($\Gamma=2.41^{+0.15}_{-0.14}$) is
NuSTAR~J121027+3929.1, the HBL previously identified at $<10$~keV
(e.g.,\ Gioia et~al. 1990; Morris et~al. 1991). Indeed, steep X-ray
spectral slopes are typical of HBLs (e.g.,\ Sambruna et~al. 1996;
Fossati et~al. 1997).

Four of the ten sources ($\approx40^{+32}_{-19}$\%) require the
presence of absorption, with $N_{\rm H}\simgt10^{22}$~cm$^{-2}$, and
the other six sources have absorbing column density upper limits. The
fraction of X-ray absorbed AGNs with $N_{\rm H}>10^{22}$~cm$^{-2}$ in
the {\it Swift}-BAT sample of Burlon et~al. (2011) is
$\approx53^{+4}_{-4}$\%, indicating no significant difference in the
fraction of absorbed AGNs between the {\it NuSTAR} sources and the
{\it Swift}-BAT AGNs. Eight of the {\it NuSTAR} sources are quasars
with $L_{\rm 10-40 keV}>10^{44}$~erg~s$^{-1}$, and four
($\approx50^{+40}_{-24}$\%) of the quasars are absorbed with $N_{\rm
  H}\simgt10^{22}$~cm$^{-2}$; see Fig.~9. The fraction of obscured
quasars is in broad agreement with that found at $\simgt10$~keV in the
local Universe and from {\it Chandra} and {\it XMM-Newton} surveys at
higher redshift (e.g.,\ Ueda et~al. 2003; La Franca et~al. 2005;
Akylas et~al. 2006; Hasinger 2008; Burlon et~al. 2011; Malizia
et~al. 2012); however, better source statistics are required to
provide sufficient constraints to distinguish between different X-ray
background synthesis models (Gilli et~al. 2007). Two of the X-ray
absorbed quasars are BLAGNs and two are NLAGNs and we discuss the
origin of the obscuration towards these sources in \S4.4.

%
%
\begin{figure}[!t]
{\centerline{\includegraphics[angle=0,width=8.0cm]{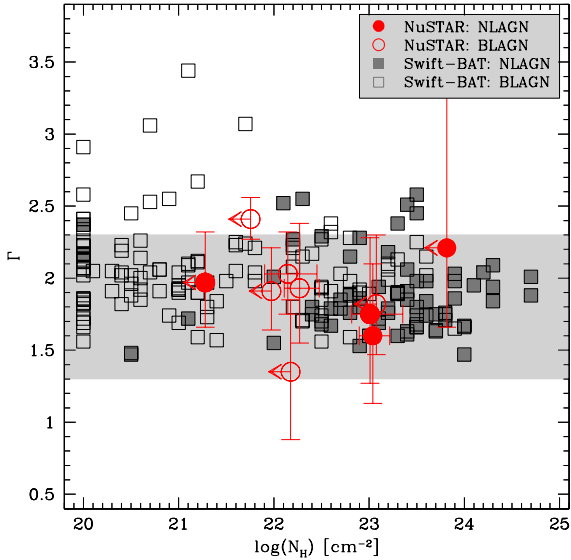}}}
\caption{Best-fitting X-ray spectral parameters ($\Gamma$ versus
  $N_{\rm H}$) for the {\it NuSTAR} sources and the {\it Swift}-BAT
  AGNs in Burlon et~al. (2011). See Fig.~6 for a description of the
  symbols; the error bars indicate 90\% confidence uncertainties for
  one interesting parameter. The shaded region indicates the range of
  properties found for local AGNs (see \S4.2).}
\end{figure}

None of the {\it NuSTAR} sources appear to be absorbed by
Compton-thick material ($N_{\rm H}\simgt10^{24}$~cm$^{-2}$), despite
the near obscuration-independent AGN selection over the {\it NuSTAR}
energy range. However, the absorbing column densities of Compton-thick
AGNs are so high that even the $>10$~keV emission can be significantly
absorbed (e.g.,\ AGNs with $N_{\rm H}\simgt5\times10^{24}$~cm$^{-2}$
can be suppressed by an order of magnitude; see Fig.~11 of Burlon
et~al. 2011). Therefore, Compton-thick AGNs can be comparatively rare
even in high-energy AGN samples.\footnote{Less direct approaches are
  often required to identify Compton-thick AGNs with $N_{\rm
    H}\simgt3\times10^{24}$~cm$^{-2}$ (e.g.,\ optical--mid-infrared
  spectroscopy, photometry, and SED fitting; Risaliti et~al. 1999;
  Alexander et~al. 2008; Treister et~al. 2009; Goulding et~al. 2011;
  Del Moro et~al. 2013; Luo et~al. 2013).} Indeed, on the basis of the
results obtained for local AGNs at $>10$~keV with the {\it INTEGRAL}
and {\it Swift}-BAT observatories, only $\approx$~5--10\% of the
detected sources are Compton-thick AGNs (e.g.,\ Tueller et~al. 2008;
Beckmann et~al. 2009; Burlon et~al. 2011; Ajello et~al. 2012), despite
the intrinsic fraction of Compton-thick AGNs likely being
substantially larger.\footnote{Assuming that the intrinsic
  distribution of absorbing column densities over $N_{\rm
    H}=10^{22}$--$10^{26}$~cm$^{-2}$ is flat (e.g.,\ Risaliti
  et~al. 1999) and that $>10$~keV surveys are only sensitive to the
  identification of AGNs with $N_{\rm
    H}\simlt3\times10^{24}$~cm$^{-2}$, the intrinsic fraction of
  Compton-thick AGNs would be $\approx$~20--40\%.} If distant AGNs
have a similar range of absorbing column densities as those found
locally, we would therefore expect $\approx$~0.5--1 Compton-thick AGNs
in our small sample; given the tentative evidence for an increase in
the fraction of obscured AGNs with redshift (e.g.,\ La Franca
et~al. 2005; Ballantyne et~al. 2006, Treister \& Urry 2006; Brightman
\& Ueda 2012), we may expect the Compton-thick AGN fraction to be even
larger in the distant Universe. Taking account of the low number
statistics of our sample, we can therefore place an upper limit to the
fraction of Compton-thick AGNs of $\simlt23$\% in our sample (90\%
confidence; see Table 1 of Gehrels 1986). The 90\% upper limit on the
fraction of Compton-thick quasars over the redshift range of
$z=$~0.5--1.1 is $\simlt33$\% if we only consider the seven {\it
  NuSTAR} sources with $L_{\rm 10-40 keV}>10^{44}$~erg~s$^{-1}$. These
upper limits are marginally too high to distinguish between different
model predictions for the fraction of Compton-thick AGNs detected in
{\it NuSTAR} surveys for a range of AGN luminosity functions and
column-density distributions ($\simlt$~22\%; Ballantyne et~al. 2011;
Akylas et~al. 2012). Better source statistics are clearly required to
accurately measure the fraction of distant Compton-thick AGNs.

\subsection{X-ray spectral properties: the presence of reflection}

A unique aspect of the {\it NuSTAR} data is the insight that it places
on the $>10$~keV emission from distant AGNs and the presence of
spectral complexity beyond that of simple power-law emission (e.g.,\ a
reflection component), particularly at $z\simlt1$ where the rest-frame
energy coverage of {\it Chandra} and {\it XMM-Newton} is comparatively
modest. By focusing on $>10$~keV emission, the effect of absorption on
the observed emission will be neglible (at least up to $N_{\rm
  H}\approx5\times10^{23}$~cm$^{-2}$) and the presence of reflection
can be revealed by the flattening of the intrinsic power-law
component.

To investigate the $>10$~keV emission in our sources we fitted the
rest-frame 10--40~keV emission using a simple power-law model (the
{\sc pow} model in {\sc xspec}), following \S3.1; see Table~4. The
spectral constraints for individual sources are poor and range from
$\Gamma_{\rm 10-40 keV}\approx$~0.4--2.4, with large uncertainties;
the mean X-ray spectral slope is $\Gamma_{\rm 10-40
  keV}\approx1.9$. However, we can place accurate average spectral
constraints by jointly fitting the data. When jointly fitting the data
we fitted the rest-frame 10--40~keV data of the {\it NuSTAR} sources
with a power-law model, jointly fitting the power-law component but
leaving the normalization for each source to vary independently. In
this analysis we excluded NuSTAR~J121027+3929.1, the HBL, and
NuSTAR~J032459-0256.1, the low-luminosity system, since we wanted to
focus on luminous non-beamed AGNs. The best-fitting X-ray spectral
slope from the joint spectral fitting is $\Gamma_{\rm 10-40
  keV}=1.88^{+0.26}_{-0.25}$, in good agreement with the intrinsic
X-ray spectral slope found for nearby AGNs studied at $>10$~keV
(e.g.,\ Deluit \& Courvoisier 2003; Dadina 2008; Molina et~al. 2009;
Burlon et~al. 2011); see Table~5. To first order, the comparatively
steep average rest-frame 10--40~keV spectral slope suggests that there
is not a significant reflection component in these sources, on
average, which would manifest itself as a relatively flat X-ray
spectral slope at $>10$~keV (e.g.,\ Nandra \& Pounds 1994).

We can more directly constrain the average strength of the reflection
component by jointly fitting the rest-frame 10--40~keV data using the
{\sc pexrav} model in {\sc xspec} (Magdziarz \& Zdziarski
1995).\footnote{The {\sc pexrav} model calculates the expected X-ray
  continuum spectrum due to the reflection of power-law emission by
  neutral material.} Fixing the X-ray spectral slope to $\Gamma=1.8$
and adopting the default parameters for {\sc pexrav} we constrain the
average strength of the reflection for the eight {\it NuSTAR} sources
to be $R<1.4$.\footnote{The reflection parameter $R$ indicates the
  solid angle of a neutral slab of material illuminated by the primary
  X-ray source: $R\approx{\Omega\over{2\pi}}$.} Conversely, if we fix
$R=1$, the typical value found for nearby AGNs selected at $>10$~keV
(e.g.,\ Deluit \& Courvoisier 2003; Dadina 2008; Beckmann et~al. 2009;
Molina et~al. 2009), we constrain the intrinsic X-ray spectral slope
to be $\Gamma=2.08^{+0.25}_{-0.24}$, also consistent with that of
nearby AGNs; see Table~5. To first order, our results therefore
suggest that the strength of reflection in distant luminous AGNs is
consistent to that found for local AGNs. However, better source
statistics are required to more accurately constrain the strength of a
reflection component in distant AGNs and to search for changes in the
reflection component within sub populations (e.g.,\ dividing the
samples in terms of luminosity and absorbing column density).

%
%
\begin{figure}[!t]
{\centerline{\includegraphics[angle=0,width=8.0cm]{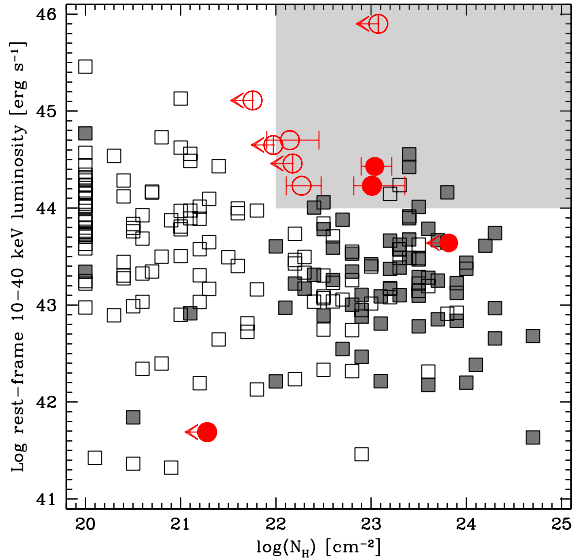}}}
\caption{Luminosity versus best-fitting absorbing column density
  ($N_{\rm H}$) for the {\it NuSTAR}-detected sources and the {\it
    Swift}-BAT AGNs in Burlon et~al. (2011). See Fig.~6 for a
  description of the symbols; the error bars indicate 90\% confidence
  uncertainties for one interesting parameter. The shaded region
  indicates the source properties expected for obscured quasars
  ($L_{\rm 10-40 keV}\simgt10^{44}$~erg~s$^{-1}$ and $N_{\rm
    H}\simgt10^{22}$~cm$^{-2}$).}
\end{figure}

\subsection{Ultraviolet--mid-infrared source properties}

The UV--MIR data of the {\it NuSTAR} sources can provide insight into
the emission from the AGN and host galaxy and the presence of dust
reddening. Below we first explore the MIR colors of the {\it NuSTAR}
sources and we then analyze their UV--MIR SEDs.

\subsubsection{Infrared color analysis}

Various work over the past decade has shown that MIR colors provide a
powerful method to robustly select luminous AGNs in a manner that is
relatively unbiased by obscuration (e.g.,\ Stern et~al. 2005, 2012;
Assef et~al. 2010, 2013; Donley et~al. 2007, 2012). As such, MIR
selection has some similarity to hard X-ray selection, and MIR and
hard X-ray source selection are potentially the two most promising
avenues for uncovering the full census of AGN in universe. Each
wavelength has various strengths and weaknesses. In particular,
various work has shown that MIR selection preferentially identifies
the most luminous AGN with quasar-level luminosities (e.g.,\ Donley
et~al. 2007, Eckart et~al. 2010), while X-ray selection efficiently
identifies moderate--high luminosity AGNs (e.g.,\ Barger et~al. 2003;
Szokoly et~al. 2004; Xue et~al. 2011). On the other hand, MIR surveys
have now mapped the entire celestial sphere, identifying millions of
robust AGN candidates. In contrast, {\it NuSTAR} is unlikely to map
more than $\approx$~10--20 deg$^2$ over its entire mission
lifetime. In order to explore this MIR--X-ray complimentarity in the
new regime offered by {\it NuSTAR}, we therefore briefly discuss the
MIR colors of the ten serendipitous {\it NuSTAR} sources.

Only one of the ten {\it NuSTAR} sources (NuSTAR~J063358+1742.4) has
four-band {\it Spitzer}-IRAC detections, a requirement for the {\it
  Spitzer} MIR AGN selection criteria; NuSTAR~J063358+1742.4 is
fainter than the {\it WISE} flux limits but has IRAC colors that place
it within the IRAC AGN wedge of Stern~et al. (2005). Of the other nine
{\it NuSTAR} sources, eight have at least two-band detections by {\it
  WISE}. Stern et~al. (2012) and Assef et~al. (2013) have recently
developed {\it WISE} AGN selection criteria, effectively extending the
{\it Spitzer} selection criteria across the full sky (see also Mateos
et~al. 2012; Wu et~al. 2012). Five of the eight {\it NuSTAR} sources
have {\it WISE} colors indicative of an AGN according to those
criteria. The outliers include the two sources with the weakest AGN
component (i.e.,\ lowest $\hat{a}$ values; see \S3.2),
NuSTAR~J011042-4604.2 and NuSTAR~JJ032459-0256.1. These are the only
sources with $\hat{a} < 0.5$, confirming that MIR selection misses
sources where the AGN is not bolometrically dominant.

The final outlier is the HBL NuSTAR~J121027+3929.1, a BL Lac-type
blazar.  Massaro et~al. (2011) have recently published a series of
papers discussing the {\it WISE} colors of blazars. While
Flat-Spectrum Radio Quasars (FSRQ) type blazars have colors typical of
other AGN populations (e.g.,\ Yan et~al. 2013), BL Lac-type blazars
have unique colors. However, as NuSTAR~J121027+3929.1 is only detected
in the two shorter wavelength bandpasses of {\it WISE}, it is not
possible to compare this source to the color criteria developed by
Massaro et~al. (2011) and Yan et~al. (2013); note also the caveat
emptor in Footnote 3 of Stern \& Assef (2013).

\subsubsection{Spectral energy distribution analysis}

To quantify the UV--MIR emission of the {\it NuSTAR} sources we fitted
the broad-band SEDs following \S3.2; see Fig.~5 and Table~3. A
significant AGN component ($\hat{a}>0.4$) is required to explain the
UV--MIR emission for all of the sources except for the low-luminosity
system NuSTAR~J032459-0256.1. The rest-frame 6~$\mu$m luminosities of
the {\it NuSTAR} sources (${\nu}L_{\rm 6~\mu
  m}\approx$~(0.9--30)~$\times10^{44}$~erg~s$^{-1}$, with the
exception of NuSTAR~J032459-0256.1, which has ${\nu}L_{\rm 6~\mu
  m}\approx4\times10^{40}$~erg~s$^{-1}$) are in general agreement with
that expected for the MIR--X-ray (i.e.,\ 6~$\mu$m--2--10~keV)
luminosity relationship found for AGNs (e.g.,\ Lutz et~al. 2004; Fiore
et~al. 2009); we assumed $\Gamma=1.8$ to convert between rest-frame
2--10~keV and rest-frame 10--40~keV. However, we note that the HBL
NuSTAR~J121027+3929.1 and the highest-redshift source
NuSTAR~J115746+6004.9 are both X-ray bright compared to the strength
of the AGN at 6~$\mu$m, suggesting that the X-ray emission from these
sources is probably beamed (as would be, at least, expected for an
HBL).

In some cases the presence of dust reddening in the best-fitting SED
solutions means that the observed contribution of the AGN at
UV--optical wavelengths is negligible. However, we highlight here
that, although the strength of the AGN continuum at UV--optical
wavelengths plotted in Fig.~5 is inconsistent with the optical
spectroscopy in some cases (e.g.,\ NuSTAR~J011042-4604.2 and
NuSTAR~J181428+3410.8), they are broadly consistent when the range in
dust reddening from the best-fitting solution is taken into account;
see Table~3. As expected on the basis of the simplest unified AGN
model (e.g.,\ Antonucci 1993), the optical emission is heavily
extinguished in the NLAGNs ($E(B-V)$~$\approx$~3--6~mags, which
corresponds to $A_{\rm V}\approx$~9--18~mags for $R_{\rm V}=3.1$;
e.g.,\ Savage \& Mathis 1979), with the exception of the
low-luminosity system NuSTAR~J032459-0256.1. There is evidence of
dust-reddening for two of the BLAGNs (NuSTAR~J181428+3410.8 has
$E(B-V)$~$\approx$~2~mags and NuSTAR~J183443+3237.8 has
$E(B-V)$~$\approx$~0.6~mags) and, as we discuss in the Appendix, the
reddening towards NuSTAR~J183443+3237.8 appears to be variable. None
of the other BLAGNs show evidence for significant obscuration at
optical wavelengths, as expected for the simplest version of the
unified AGN model for BLAGNs; however, we note that there is a large
uncertainty in the dust reddening for NuSTAR~J011042-4604.2, which is
due to the limited number of photometric data points.

%
%
\begin{figure}[!t]
\includegraphics[angle=0,width=8.0cm]{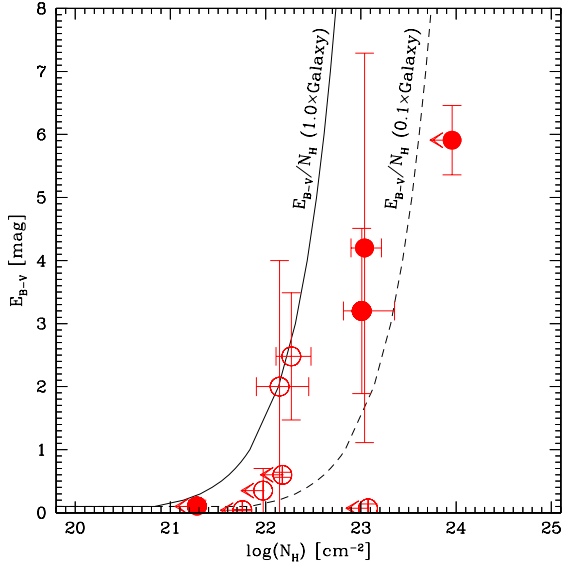}
\caption{Dust reddening ($E(B-V)$) versus X-ray absorption ($N_{\rm
    H}$) for the {\it NuSTAR} serendipitous sources from the UV--MIR
  SED fitting and the X-ray spectral fitting, respectively; see Fig.~6
  for the description of the symbols. The solid curve indicates the
  relationship between dust reddening and X-ray absorption expected
  from the $A_{\rm V,Gal}$--$N_{\rm H,Gal}$ relationship found in the
  Galaxy (Savage \& Mathis 1979; G{\"u}ver \& {\"O}zel 2009) while the
  dashed curve indicates $A_{\rm V,Gal}$--$10\times{N_{\rm H,Gal}}$,
  broadly consistent with that found for AGNs by Maiolino
  et~al. (2001).}
\end{figure}

In Fig.~10 we compare the obscuration estimated from the UV--MIR SED
fitting to that measured from the X-ray spectral fitting, which
provides constraints on the dust-to-gas ratio in AGNs. The BLAGN with
the strongest dust reddening (NuSTAR~J181428+3410.8) has a measured
X-ray absorbing column density of $N_{\rm H}\approx10^{22}$~cm$^{-2}$,
consistent with that expected given the $A_{\rm V}$--$N_{\rm H}$
relationship found in the Galaxy (e.g.,\ G{\"u}ver \& {\"O}zel
2009). The constraints on the X-ray absorbing column density for the
other four BLAGNs are also consistent with that expected given the
$A_{\rm V}$--$N_{\rm H}$ relationship found in the Galaxy; however, in
all cases the column-density constraints are too weak to rule out the
different $A_{\rm V}$--$N_{\rm H}$ relationship found by Maiolino
et~al. (2001). By comparison, although the three NLAGNs with $L_{\rm
  10-40~keV}\simgt10^{43}$~erg~s$^{-1}$ (NuSTAR~J063358+1742.4,
NuSTAR~J115912+4232.6, and NuSTAR~J204021-0056.1) have evidence for
significant obscuration, the inferred X-ray absorbing column density
from the dust reddening measurements are lower than those directly
measured from the X-ray spectral analyzes ($N_{\rm
  H,A_{V}}\approx$~(2--4)~$\times10^{22}$~cm$^{-2}$, as compared to
$N_{\rm H}\approx10^{23}$~cm$^{-2}$ measured from the X-ray data).
However, the dust-to-gas ratios are consistent with the lower $A_{\rm
  V}$--$N_{\rm H}$ relationship found by Maiolino et~al. (2001) for
AGNs. Neither the HBL NuSTAR~J121027+3929.1 nor the low-luminosity
system NuSTAR~J032459-0256.1 show evidence for significant obscuration
in the UV--MIR and X-ray bands.

The best-fitting SED solutions also provide a first-order estimate of
the host-galaxy stellar masses of the {\it NuSTAR} sources. The range
of stellar masses is large, from $\approx2\times10^{9}$~$M_{\odot}$
(for the low-luminosity system NuSTAR~J032459-0256.1) to
$\approx3\times10^{12}$~$M_{\odot}$ (for the highest-redshift source
NuSTAR~J115746+6004.9). However, the stellar masses for the majority
of the {\it NuSTAR} sources are relatively tightly constrained: the
stellar-mass range with these two extreme sources removed is
(0.7--3.3)~$\times10^{11}$~$M_{\odot}$, and the median stellar mass is
$\approx10^{11}$~$M_{\odot}$. Many of the {\it NuSTAR} sources are
BLAGNs and we caution that reliable stellar-mass constraints are
challenging for these systems due to the contribution of the AGN to
the rest-frame optical--near-IR emission (see \S3.2 and Fig.~5 for the
SED-fitting constraints). However, reassuringly, the median stellar
mass of the NLAGNs, where accurate stellar-mass constraints are less
challenging, is consistent with that of the BLAGNs when the two
extreme sources are removed ($\approx10^{11}$~$M_{\odot}$).

The range and median stellar mass of the {\it NuSTAR} sources are
similar to those of comparably distant AGNs detected at $<10$~keV in
{\it Chandra} and {\it XMM-Newton} surveys (e.g.,\ Babi{\'c}
et~al. 2007; Alonso-Herrero et~al. 2008; Bundy et~al. 2008; Xue
et~al. 2010; Lusso et~al. 2011). However, by comparison, the median
stellar mass of the {\it NuSTAR} sources is $\approx$~5 times higher
than for $z<0.05$ AGNs detected at $>10$~keV by {\it Swift}-BAT
($\approx2\times10^{10}$~$M_{\odot}$; Koss et~al. 2011). To first
order this suggests that there has been significant evolution in the
characteristic mass of high-energy emitting AGNs over the redshift
range $z\approx$~0--1. However, the {\it NuSTAR} sources are more
luminous than the {\it Swift}-BAT AGNs and that could bias the results
towards more massive systems. For example, for a constant average
Eddington ratio, the order of magnitude higher median X-ray luminosity
of the {\it NuSTAR} sources over the {\it Swift}-BAT AGNs (see \S4.1)
would lead to an order of magnitude higher black hole mass and thereby
a larger stellar mass, assuming no evolution in the
black-hole--spheroid mass relationship (e.g.,\ Magorrian et~al. 1998;
Marconi \& Hunt 2003; G{\"u}ltekin et~al. 2009). Indeed, Koss
et~al. (2011) show a weak trend between mean stellar mass and X-ray
luminosity for the {\it Swift}-BAT AGNs. Therefore, while our results
indicate that the most luminous high-energy emitting AGNs at
$z\simgt$~0.1 are hosted by more massive galaxies than high-energy
emitting AGNs at $z<0.05$, a systematic analysis of both local and
distant AGNs taking account of potential X-ray luminosity biases, is
required to derive more accurate constraints.

%
\section{Conclusions}
%

We have reported on the first ten identifications of {\it NuSTAR}
sources serendipitously detected in the extragalactic survey
programme. These {\it NuSTAR} sources are $\approx$~100 times fainter
than AGNs previously detected at $>10$~keV and have a broad range in
redshift and luminosity ($z=$~0.020--2.923 and $L_{\rm 10-40
  keV}\approx4\times10^{41}$--$5\times10^{45}$~erg~s$^{-1}$); the
median redshift and luminosity are $z\approx$~0.7 and $L_{\rm 10-40
  keV}\approx3\times10^{44}$~erg~s$^{-1}$, respectively. On the basis
of broad-band $\approx$~0.5--32~keV spectroscopy, optical
spectroscopy, and broad-band UV--MIR SED analyzes we found the
following results:

\begin{itemize}

\item five ($\approx50^{+34}_{-22}$\%) of the ten {\it NuSTAR} sources
  are classified as broad-line AGNs (BLAGNs), four
  ($\approx40^{+32}_{-19}$\%) are classified as narrow-line AGNs
  (NLAGNs), and one is a BL Lac. The BLAGN:NLAGN ratio is consistent
  with that found for $\simgt$~10~keV selected AGNs in the local
  Universe. See \S4.1.

\item from fitting the broad-band X-ray spectra we find that the
  dominant source population are quasars with $L_{\rm 10-40
    keV}>10^{44}$~erg~s$^{-1}$, of which $\approx$~50\% are obscured
  with $N_{\rm H}\simgt10^{22}$~cm$^{-2}$. However, none of the seven
  quasars over the redshift range $z=$~0.5--1.1 are Compton thick and
  we place a 90\% confidence upper limit of $\simlt$~33\% on the
  Compton-thick quasar fraction. See \S4.2.

\item from jointly fitting the rest-frame $\approx$~10--40~keV data
  for all of the non-beamed sources with $L_{\rm 10-40
    keV}>10^{43}$~erg~s$^{-1}$ we constrain the high-energy X-ray
  spectral slope and the average strength of a reflection
  component. We find $R<1.4$ for $\Gamma=1.8$ and
  $\Gamma=2.08^{+0.25}_{-0.24}$ and $R=1.0$, consistent with that
  found for local AGNs selected at $>10$~keV. See \S4.3.

\item from fitting the UV--MIR SEDs we constrain the stellar masses of
  the host galaxies, finding a median stellar mass of
  $\approx10^{11}$~$M_{\odot}$. The host galaxies of {\it NuSTAR}
  sources are $\approx$~5 times more massive on average than {\it
    Swift}-BAT-detected local AGNs at $>10$~keV. At least part of this
  implied evolution in the characteristic mass of high-energy emitting
  AGNs is likely to be due to X-ray luminosity biases. See \S4.4.

\end{itemize}

{\it NuSTAR} is providing unique insight into the high-energy
properties of AGNs, achieving a factor $\approx$~100 times improvement
in sensitivity over previous observatories at $\simgt10$~keV. In the
current study we do not find significant differences in the fraction
of absorbed AGNs between the {\it NuSTAR} sources and nearby
high-energy emitting AGNs, despite the {\it NuSTAR} sources being
$\approx$~10 times more luminous (and $\approx$~5 times more massive),
on average. These results therefore suggest that the central engine of
distant high-energy emitting AGNs are similar to that of nearby
AGNs. However, the current study is limited in source statistics and
provides a first look at the high-energy properties of distant
AGNs. With the $\approx$~20--40 times improvement in sample size
afford by the full {\it NuSTAR} extragalactic survey (completed in the
first 2 years of {\it NuSTAR} observations) we will be able to make
more detailed comparisons and accurately measure the high-energy
properties of distant AGNs and constrain their evolution with
redshift.

\acknowledgements We acknowledge financial support from the Leverhulme
Trust (DMA; JRM), the Science and Technology Facilities Council (STFC;
DMA; ADM; GBL), the SAO grant GO2-13164X (MA), NASA Postdoctoral
Program at the Jet Propulsion Laboratory (RJA), NSF award AST 1008067
(DRB), Center of Excellence in Astrophysics and Associated
Technologies (PFB~06/2007; FEB; ET), the Anillo project ACT1101 (FEB;
ET), FONDECYT Regular 1101024 (FEB), Caltech NuSTAR subcontract
44A-1092750 (WNB; BL), NASA ADP grant NNX10AC99G (WNB; BL), ASI/INAF
grant I/037/12/0 (AC; SP), CONICYT-Chile under grant FONDECYT 3120198
(CS), and FONDECYT regular grant 1120061 (ET). We thank the referee
for a constructive and positive report. We also thank Michael Koss for
the discussion of {\it Swift}-BAT results, and Mark Brodwin, Daniel
Gettings, John Gizis, Richard Walters, Jingwen Wu, and Dominika
Wylezalek for supporting the ground-based follow-up observations. This
work was supported under NASA Contract No. NNG08FD60C, and made use of
data from the {\it NuSTAR} mission, a project led by the California
Institute of Technology, managed by the Jet Propulsion Laboratory, and
funded by the National Aeronautics and Space Administration. We thank
the {\it NuSTAR} Operations, Software and Calibration teams for
support with the execution and analysis of these observations.  This
research has made use of the {\it NuSTAR} Data Analysis Software
(NuSTARDAS) jointly developed by the ASI Science Data Center (ASDC,
Italy) and the California Institute of Technology (USA).



%
%

%
%

\clearpage
\LongTables 
\begin{landscape}
\begin{deluxetable}{lcccccccccc}
\tabletypesize{\footnotesize}
\tablewidth{0pt}
\tablecaption{X-ray observations used in the paper}
\tablehead{
\colhead{Target Field} &
\colhead{HLX~1} &
\colhead{NGC~1320} &
\colhead{Geminga}  &
\colhead{SDSS~J1157+6003} &
\colhead{IC~751} &
\colhead{NGC~4151} &
\colhead{Cen~X-4} &
\colhead{WISE~J1814+3412} &
\colhead{3C~382} &
\colhead{AE~Aqr}}
\startdata
Observatory  & {\it NuSTAR}    & {\it NuSTAR}    & {\it NuSTAR}      & {\it NuSTAR}    & {\it NuSTAR} & {\it NuSTAR}    & {\it NuSTAR}    & {\it NuSTAR}    & {\it NuSTAR}    & {\it NuSTAR} \\
Observation  & 30001030002     & 60061036002     & 30001029(002--028)$^b$ & 60001071002     & 60061217004  & 60001111005     & 30001004002     & 6000111402      & 60061286002     & 30001120004  \\
Start date         & 2012-11-19      & 2012-10-25      & 2012-09-26        & 2012-10-28      & 2013-02-04   & 2012-11-14      & 2013-01-20      & 2012-10-30      & 2012-09-18      & 2012-09-05   \\
Exposure$^a$ & 177.1~ks        & 14.5~ks         & 142.6~ks          & 21.7~ks         & 56.1~ks      & 61.8~ks         & 116.4~ks        & 21.3~ks         & 16.6~ks         & 71.3~ks      \\
             &                 &                 &                   &                 &                 &              &                 &                 &                 & \\
Observatory  & {\it Swift}-XRT & {\it Swift}-XRT & {\it Chandra}     & {\it Chandra}   & \nodata      & {\it Swift}-XRT & {\it XMM-Newton}& {\it XMM-Newton}& {\it Swift}-XRT & {\it XMM-Newton} \\
Observation  & 00031287003     & 00080314001     & 7592              & 5698            & \nodata      & 00080073001     & 0144900101      & 0693750101      & 00080217001     & 0111180201    \\
Start date         & 2008-11-07      & 2012-10-26      & 2007-08-27        & 2005-06-03      & \nodata      & 2012-11-20      & 2003-03-01      & 2012-10-07      & 2012-09-18      & 2001-11-07    \\
Exposure$^a$ & 11.3~ks         & 6.8~ks          & 77.1~ks           & 7.0~ks          & \nodata      & 1.1~ks          & 55.3~ks         & 29.6~ks         & 6.6~ks          & 4.3~ks        \\
\enddata
\tablecomments{$^a$~the nominal on-axis exposure time (for {\it
    NuSTAR} the exposure is from FPMA), corrected for background
  flaring and bad events; $^b$~the range of observation numbers that
  have been combined to produce the final image (only the 15
  observations ending in even numbers are used).}
\end{deluxetable}
\clearpage
\end{landscape}

%
%

\clearpage
\LongTables 
\begin{landscape}
\begin{deluxetable}{lcccccccccc}
\tabletypesize{\scriptsize}
\tablewidth{0pt}
\tablecaption{NuSTAR Source Properties}
\tablehead{
\colhead{Target Field} &
\colhead{HLX~1} &
\colhead{NGC~1320} &
\colhead{Geminga}  &
\colhead{SDSS~J1157+6003} &
\colhead{IC~751} &
\colhead{NGC~4151} &
\colhead{Cen~X4} &
\colhead{WISE~J1814+3412} &
\colhead{3C~382} &
\colhead{AE~Aqr}\\
\colhead{Source Name$^a$} &
\colhead{011042-4604.2} &
\colhead{032459-0256.1} &
\colhead{063358+1742.4}  &
\colhead{115746+6004.9} &
\colhead{115912+4232.6} &
\colhead{121027+3929.1} &
\colhead{145856-3135.5} &
\colhead{181428+3410.8} &
\colhead{183443+3237.8} &
\colhead{204021-0056.1}}
\startdata
RA~(J2000)$^b$      & 01:10:42.7    & 03:24:59.5    & 06:33:58.2   & 11:57:46.2     & 11:59:12.4   & 12:10:27.0    & 14:58:56.6    & 18:14:28.2    & 18:34:43.6    & 20:40:21.0    \\
DEC~(J2000)$^b$     & $-$46:04:17    & $-$02:56:09  & +17:42:25   & +60:04:55     & +42:32:37   & +39:29:07    & $-$31:35:34  & +34:10:51    & +32:37:52    & $-$00:56:06  \\
                   &                &                &               &                 &               &                &                &                &                &    \\
Exposure (A)$^c$   & 159.8          & 8.0            & 92.4          & 18.2            & 31.0          & 42.1           &  27.4          &  18.4          & 9.2            &  52.4  \\
Exposure (B)$^c$   & 159.4          & 5.9            & 93.0          & 18.5            & 31.2          & 39.9           &  46.0          &  20.0          & 9.6            &  51.7  \\
                   &                &                &               &                 &               &                &                &                &                &    \\
3--24~keV (A)$^d$  & $295\pm35$ & $129\pm17$ & $102\pm24$& ($31\pm12$) & $213\pm20$& $621\pm34$ & $34\pm22$   & $<28$       & $43\pm12$  & $148\pm27$ \\
3--8~keV  (A)$^d$  & $172\pm25$ & $90\pm13$  & $51\pm16$ & ($24\pm9$)  & $132\pm14$& $477\pm28$ & $<33$       & $<19$       & ($20\pm8$) & $92\pm19$  \\
8--24~keV (A)$^d$  & $123\pm25$ & $38\pm11$  & $52\pm18$ & $<18$       & $82\pm14$ & $145\pm20$ & $<36$       & ($16\pm9$)  & ($24\pm8$) & $57\pm19$  \\
3--24~keV (B)$^d$  & $265\pm37$ & $97\pm14$  &($87\pm31$)& $35\pm12$   & $262\pm24$& $655\pm33$ & $28\pm19$   & $23\pm12$   & $52\pm11$  & $107\pm24$  \\
3--8~keV  (B)$^d$  & $158\pm26$ & $62\pm10$  & $59\pm22$ & $19\pm8$    & $156\pm17$& $494\pm27$ & $<25$       & ($14\pm8$)  & $30\pm8$   & $77\pm18$ \\
8--24~keV (B)$^d$  & $109\pm26$ & $35\pm10$  & $<50$     & ($16\pm8$)  & $108\pm17$& $172\pm19$ & $29\pm15$   & ($9\pm9$)   & ($22\pm7$) & $<37$ \\
Aperture$^e$       &   45           &  60            &  45           &  45             &  45           &  30            &  45            &  45            &  45            &  45\\
                   &                &                &               &                 &               &                &                &                &                &    \\
Flux (3--24~keV)$^f$ &  1.3        & 9.2           & 0.8          & 1.2$^h$        & 5.6          & 12.1          &   0.6$^h$     &  0.9$^h$      & 3.5$^h$       &  1.6      \\
Flux (3--8~keV)$^f$  &  0.4        & 3.9           & 0.3          & 0.5$^h$        & 1.9          & 6.7           &  $<$0.2$^h$   &  0.3$^h$      & 1.1$^h$       &  0.7      \\
Flux (8--24~keV)$^f$ &  0.8        & 5.8           & 0.6          & 0.7$^h$        & 4.1          & 5.9           &   0.9$^h$     &  0.7$^h$      & 3.2$^h$       &  1.0      \\
               &                &                &               &                 &               &                &                &                &                &    \\
X-ray offset$^g$   &  4.3           & 7.6            & 0.9           & 4.6             &  \nodata      & 4.9            & 6.5            &  9.9           &  8.6           &  4.5   \\
Flux (3--8~keV; other data)$^i$  &  0.8        & 2.7           & 0.2          & 0.8            &  \nodata      & 7.5           & 0.4           &  0.4          & 1.5           &  0.7      \\
\enddata
\tablecomments{$^a$ source name (NuSTAR~J), based on the
  counts-weighted {\it NuSTAR} source position following the IAU
  source-name convention (see Footnote 3); $^b$ counts-weighted {\it
    NuSTAR} source position measured in the 3--24~keV energy band (see
  \S2.1.4); $^c$ effective exposure at the source position in FPMA and
  FPMB in units of ks. The effective exposure is measured from the
  exposure maps (see \S2.1.1); $^d$ net counts, 1~$\sigma$
  uncertainties, and 3~$\sigma$ upper limits measured at the
  counts-weighted {\it NuSTAR} source position in the 3--24~keV,
  3--8~keV, and 8--24~keV bands for FPMA and FPMB (see \S2.1.2). The
  values in parentheses indicate a lower significance counterpart (see
  \S2.1.1); $^e$ radius (in arcseconds) of the circular aperture used
  to measure the source photometry (see \S2.1.2); $^f$
  aperture-corrected flux in the 3--24~keV, 3--8~keV, and 8--24~keV
  energy bands in units of $10^{-13}$~erg~s$^{-1}$~cm$^{-2}$ (see
  \S2.1.3); $^g$ positional offset (in arcseconds) between the
  counts-weighted {\it NuSTAR} source position and the closest source
  detected in the lower-energy X-ray data (i.e.,\ {\it Chandra}, {\it
    Swift}-XRT, {\it XMM-Newton}). See Table~1; $^h$ low-count source
  and $\Gamma=1.8$ is used to convert the {\it NuSTAR} count rates
  into fluxes; $^i$ flux at 3--8~keV measured from the lower-energy
  X-ray data (either {\it Chandra}, {\it Swift}-XRT, or {\it
    XMM-Newton; see Table~1}) in units of
  $10^{-13}$~erg~s$^{-1}$~cm$^{-2}$ (see \S2.2).}
\end{deluxetable}
\clearpage
\end{landscape}

%
%

\clearpage
\LongTables 
\begin{landscape}
\begin{deluxetable}{lcccccccccc}
\tabletypesize{\scriptsize}
\tablewidth{0pt}
\tablecaption{Ultraviolet to mid-infrared source properties}
\tablehead{
\colhead{Target Field} &
\colhead{HLX~1} &
\colhead{NGC~1320} &
\colhead{Geminga} &
\colhead{SDSS~J1157+6003} &
\colhead{IC~751} &
\colhead{NGC~4151} &
\colhead{Cen~X4} &
\colhead{WISE~J1814+3412} &
\colhead{3C~382} &
\colhead{AE~Aqr}\\
\colhead{Source Name$^a$} &
\colhead{011042-4604.2} &
\colhead{032459-0256.1} &
\colhead{063358+1742.4}  &
\colhead{115746+6004.9} &
\colhead{115912+4232.6} &
\colhead{121027+3929.1} &
\colhead{145856-3135.5} &
\colhead{181428+3410.8} &
\colhead{183443+3237.8} &
\colhead{204021-0056.1}}
\startdata
RA~(J2000)$^b$    & 01:10:43.08     & 03:24:59.95     & 06:33:58.22     & 11:57:46.75     & 11:59:12.20     & 12:10:26.61     & 14:58:57.05     & 18:14:28.82     & 18:34:43.23     & 20:40:20.71  \\
DEC~(J2000)$^b$   & $-$46:04:20.0   & $-$02:56:12.1   & +17:42:24.2     & +60:04:52.9     & +42:32:35.4     & +39:29:08.4     & $-$31:35:37.8   & +34:10:51.2     & +32:37:54.4     & $-$00:56:06.0 \\
             &                 &                 &                 &                 &                 &                 &                 &                 &                 & \\
Optical offset$^c$ &  4.7 (0.8)     &  7.7 (0.5)      &  1.4 (1.1)      &  4.5 (0.1)      &  2.5           &  4.5 (0.5)      &  7.4 (1.3)      &  8.1 (1.9)           &  6.0 (2.9)      &  4.3 (0.2) \\
             &                 &                 &                 &                 &                 &                 &                 &                 &                 & \\
FUV$^d$          & \nodata         & $19.49\pm0.12$  & \nodata         & \nodata         & \nodata         & $21.81\pm0.45$  & \nodata         & \nodata         & $23.33\pm0.29$  & \nodata \\
NUV$^d$          & $21.92\pm0.45$    & $19.02\pm0.06$  & \nodata         & \nodata         & $22.31\pm0.46$  & $21.01\pm0.22$  & \nodata         & $21.07\pm0.23$  & $21.90\pm0.05$  & $23.70\pm0.25$ \\
$u$$^d$          & \nodata         & \nodata       & \nodata         & $24.5\pm1.1$    & $21.61\pm0.40$  & $20.02\pm0.04$  & \nodata         & \nodata         & \nodata         & $23.84\pm0.82$ \\
$g/B^\ddag$$^d$  & \nodata         & \nodata       & \nodata         & $22.29\pm0.13$   & $19.34\pm0.02$  & $19.72\pm0.01$  & \nodata         & $21.19\pm0.08$  & $21.63\pm0.45^\ddag$   & $22.43\pm0.14$ \\
$r/R^\ddag$$^d$  & \nodata         & 16.3$^\ddag$  & \nodata         & $21.56\pm0.09$   & $18.14\pm0.01$  & $19.42\pm0.01$  & 19.9$^\ddag$    & $20.62\pm0.07$  & $19.54\pm0.18^\ddag$   & $21.56\pm0.10$ \\
$i/I^\ddag$$^d$  & \nodata         & \nodata       & \nodata         & $21.53\pm0.13$   & $17.59\pm0.01$  & $19.06\pm0.01$  & \nodata         & \nodata         & $18.67\pm0.20^\ddag$   & $20.37\pm0.06$ \\
$z$$^d$          & \nodata         & \nodata       & \nodata         & $21.52\pm0.47$   & $17.24\pm0.03$  & $18.79\pm0.04$  & \nodata         & \nodata         & \nodata         & $20.35\pm0.22$ \\
$J$$^d$          & \nodata         & $14.89\pm0.06$  & $>19.8$         & \nodata         & $16.59\pm0.14$  & \nodata         & \nodata         & \nodata         & \nodata         & \nodata \\
$H$$^d$          & \nodata         & $14.37\pm0.09$  & \nodata         & \nodata         & $16.18\pm0.17$  & \nodata         & \nodata         & \nodata         & \nodata         & \nodata \\
$K_s$$^d$        & \nodata         & $14.06\pm0.09$  & \nodata         & \nodata         & $15.09\pm0.11$  & \nodata         & \nodata         & \nodata         & \nodata         & \nodata \\
{\it WISE} W1 (3.4~$\mu$m)$^d$           & $15.98\pm0.07$  & $13.20\pm0.03$  & \nodata         & $16.77\pm0.10$  & $14.10\pm0.03$  & $14.86\pm0.04$  & $16.91\pm0.16$  & \nodata         & $15.32\pm0.04$  & $16.05\pm0.08$ \\
{\it WISE} W2 (4.6~$\mu$m)$^d$           & $15.44\pm0.13$  & $13.00\pm0.03$  & \nodata         & $15.63\pm0.13$  & $12.97\pm0.03$  & $14.59\pm0.64$  & $15.19\pm0.12$  & \nodata         & $14.31\pm0.05$  & $15.05\pm0.10$ \\
{\it WISE} W3 (12~$\mu$m)$^d$           & $12.48\pm0.39$  & $ 9.87\pm0.05$  & \nodata         & $12.89\pm0.47$  & $ 9.70\pm0.04$  & \nodata         & \nodata         & $11.54\pm0.19$  & $11.38\pm0.11$  & $11.89\pm0.26$ \\
{\it WISE} W4 (22~$\mu$m)$^d$           & \nodata         & $ 7.70\pm0.14$  & \nodata         & \nodata         & $ 6.87\pm0.09$  & \nodata         & \nodata         & \nodata         & $ 8.62\pm0.26$  & \nodata \\
{\it Spitzer} (3.6~$\mu$m)$^d$ & \nodata & \nodata         & $ 48.78\pm0.33$ & \nodata         & $639.2\pm1.1$ & $236.79\pm0.47$ & \nodata         & $ 85.74\pm0.80$ & \nodata         & \nodata \\
{\it Spitzer} (4.5~$\mu$m)$^d$ & \nodata    & \nodata         & $ 54.13\pm0.22$ & \nodata         & $1022.07\pm0.76$ & $204.72\pm0.48$ & \nodata         & $151.22\pm0.74$ & \nodata         & \nodata \\
{\it Spitzer} (5.8~$\mu$m)$^d$ & \nodata    & \nodata         & $ 62.4\pm1.7$ & \nodata         & \nodata         & \nodata         & \nodata         & \nodata         & \nodata         & \nodata \\
{\it Spitzer} (8.0~$\mu$m)$^d$ & \nodata    & \nodata         & $114.6\pm2.0$ & \nodata         & \nodata         & \nodata         & \nodata         & \nodata         & \nodata         & \nodata \\
             &                 &                 &                 &                 &                 &                 &                 &                 &                 & \\
Redshift$^e$     & 1.073           & 0.020           & 0.891           & 2.923           & 0.177           & 0.615           & 1.045           & 0.763           & 0.510           & 0.601 \\
Telescope$^f$    & Gemini-S        & Keck            & Keck            & P200            & P200            & \nodata            & \nodata         & Keck            & P200            & Keck \\
Camera$^f$       & GMOS-S          & LRIS            & LRIS            & DBSP            & DBSP            & \nodata            & \nodata         & LRIS            & DBSP            & DEIMOS \\
UT Date$^f$      & 2012 Dec 12     & 2012 Nov 9      & 2013 Jan 10     & 2012 Nov 20     & 2012 Nov 20     & \nodata            & \nodata         & 2012 Nov 9      & 2012 Oct 10     & 2012 Oct 13 \\
Type$^f$         & BLAGN           & NLAGN           & NLAGN           & BLAGN          & NLAGN           & BL Lac          & BLAGN           & BLAGN           & BLAGN           & NLAGN\\
             &                 &                 &                 &                 &                 &                 &                 &                 &                 & \\
$\hat{a}$$^g$    & $0.43\pm0.14$   & $0.25\pm0.03$   & $0.84\pm0.04$   & $0.77\pm0.09$   & $0.70\pm0.01$   & $0.50\pm0.03$   & $1.00\pm0.04$   & $0.77\pm0.05$   & $0.67\pm0.04$ & $0.76\pm0.04$ \\
$E(B-V)$$^g$       & $2.0\pm4.6$   & $0.1\pm0.1$   & $4.2\pm2.9$   & $0.1\pm0.4$   & $5.9\pm0.6$   & $0.04\pm0.01$   & $0.0\pm0.1$   & $2.5\pm1.0$   & $0.6\pm0.5$   & $3.2\pm1.3$ \\
$L_{6\mu m}$$^g$ & $3.6\pm2.5$   & $0.004\pm0.001$   & $2.6\pm1.4$   & $29.9\pm14.3$ & $0.9\pm0.1$  & $2.1\pm0.2$   & $6.8\pm0.8$  & $4.7\pm1.1$    & $2.0\pm0.2$   & $2.2\pm0.5$ \\
$M_{*}$$^g$      & $334\pm51$    & $2.0\pm0.1$        & $114\pm25$    & $2000\pm340$  & $88\pm3$      & $236\pm18$    & $<41$       & $68\pm17$     & $117\pm20$  & $121\pm13$ \\
\tablecomments{$^a$ source name (NuSTAR~J); see Table~2; $^b$
  counterpart source position; $^c$ positional offset (in arcseconds)
  between the counts-weighted {\it NuSTAR} position and the
  counterpart source position (the value in parentheses gives the
  positional offset between the lower-energy X-ray source and the
  counterpart source position); $^d$ source photometry given in its
  native format (e.g.,\ AB mag for {\it GALEX}, AB sinh mag for SDSS,
  $\mu$Jy for {\it Spitzer}, and Vega mag for all others unless
  otherwise noted). Optical photometry with double-dagger symbol
  (\ddag) indicates when the given measurements are not from the SDSS;
  the photometry for these sources is obtained from the DSS, via
  SuperCOSMOS unless otherwise noted in the text (see \S2.4).  For the
  Geminga serendipitous source, we obtained $J$-band imaging from the
  KPNO~2.1-m telescope (see \S2.4); $^e$ optical spectroscopic
  redshift, as described in \S2.5, except for NuSTAR~J121027+3929.1
  and NuSTAR~J145856-3135.5, which are taken from Morris et~al. (1991)
  and Caccianiga et~al. (2008), respectively; $^f$ observational
  details of the optical spectroscopy and the optical spectroscopic
  classification, as given in \S2.5, \S4.1, and the Appendix (see
  Morris et~al. 1991 and Caccianiga et~al. 2008 for details of
  NuSTAR~J121027+3929.1 and NuSTAR~J145856-3135.5); $^g$ best-fitting
  parameters and 1~$\sigma$ uncertainties from the UV--mid-infrared
  SED fitting (see \S3.2): $\hat{a}$ is the fractional contribution to
  the UV--MIR emission from the AGN component, $E(B-V)$ is the dust
  reddening (units of mags), $L_{\rm 6\mu m}$ is the infrared
  luminosity of the AGN at rest-frame 6~$\mu$m ($\nu$$L_{\nu}$) in
  units of $10^{44}$~erg~s$^{-1}$, and $M_{*}$ gives the stellar mass
  (units of $10^9$~$M_{\odot}$).}
\end{deluxetable}
\clearpage
\end{landscape}

%
%

\clearpage
\LongTables 
\begin{landscape}
\begin{deluxetable}{lcccccccccc}
\tabletypesize{\scriptsize}
\tablewidth{0pt}
\tablecaption{Best-fitting Model Parameters}
\tablehead{
\colhead{Target Field} &
\colhead{HLX~1} &
\colhead{NGC~1320} &
\colhead{Geminga} &
\colhead{SDSS~J1157+6003} &
\colhead{IC~751} &
\colhead{NGC~4151} &
\colhead{Cen~X4} &
\colhead{WISE~J1814+3412} &
\colhead{3C~382} &
\colhead{AE~Aqr}\\
\colhead{Source Name$^a$} &
\colhead{011042-4604.2} &
\colhead{032459-0256.1} &
\colhead{063358+1742.4}  &
\colhead{115746+6004.9} &
\colhead{115912+4232.6} &
\colhead{121027+3929.1} &
\colhead{145856-3135.5} &
\colhead{181428+3410.8} &
\colhead{183443+3237.8} &
\colhead{204021-0056.1}}
\startdata
Data fitted$^b$    & {\it NuSTAR}          & {\it NuSTAR}          & {\it NuSTAR}          & {\it NuSTAR}          & {\it NuSTAR}          & {\it NuSTAR}          & {\it NuSTAR}          & {\it NuSTAR}          & {\it NuSTAR}          & {\it NuSTAR} \\
Energy range$^c$   & 4--32                 & 4--32                 & 4--32                 & 4--32                 & 4--32                 & 4--50                 & 4--32                 & 4--32                 & 4--32                 & 4--32  \\
$\Gamma$$^d$                 & $1.9^{+0.4}_{-0.3}$& $2.2^{+0.5}_{-0.5}$& $1.6^{+0.6}_{-0.5}$ & $2.2^{+0.8}_{-0.7}$& $1.9^{+0.3}_{-0.3}$& $2.4^{+0.2}_{-0.1}$& $0.5^{+1.2}_{-1.3}$& $1.9^{+7.4}_{-2.5}$& $1.5^{+0.7}_{-0.6}$& $1.6^{+0.5}_{-0.5}$ \\
$\Gamma_{\rm 10-40 keV}$$^e$ & $1.9^{+0.5}_{-0.5}$& $1.2^{+1.9}_{-1.2}$& $2.0^{+0.7}_{-0.6}$ & $1.9^{+0.8}_{-0.8}$& $1.8^{+1.3}_{-0.9}$& $2.4^{+0.3}_{-0.3}$& $0.4^{+1.4}_{-1.4}$& $0.4^{+2.2}_{-0.4}$& $1.7^{+1.1}_{-0.9}$& $2.3^{+0.9}_{-0.8}$ \\
$L_{\rm 10-40 keV}$$^e$      & 5.0                &  0.0049               & 2.7                & 82                & 0.37                & 13                & 4.5                & 1.7                & 2.9                & 1.7  \\
                         & & & & & & & & & & \\
Data fitted$^b$     & {\it NuSTAR}          & {\it NuSTAR}          & {\it NuSTAR}         & {\it Chandra}         & {\it NuSTAR}          & {\it NuSTAR}          & {\it XMM-Newton}      & {\it XMM-Newton}      & {\it NuSTAR}          & {\it NuSTAR} \\
                         & +{\it Swift}-XRT      & +{\it Swift}-XRT      & +{\it Chandra}        &                       &                       & +{\it Swift}-XRT      &                       &                       & +{\it Swift}-XRT      & +{\it XMM-Newton} \\
Energy range$^c$   & 0.5--32               & 0.5--32               & 0.5-32                & 0.5--8                & 4--32                 & 0.5--50               & 0.5--12               & 0.5--12               & 0.5--32               & 0.5--32 \\
$\Gamma$$^f$       & $2.0^{+0.3}_{-0.3}$& $2.0^{+0.4}_{-0.3}$& $1.6^{+0.5}_{-0.5}$& $1.9^{+0.8}_{-0.6}$& $2.2^{+1.2}_{-0.6}$& $2.4^{+0.2}_{-0.1}$& $1.9^{+0.3}_{-0.3}$& $1.9^{+0.5}_{-0.4}$& $1.4^{+0.5}_{-0.5}$& $1.8^{+0.5}_{-0.5}$ \\
$N_{\rm H}$$^f$    & $1.4^{+1.4}_{-1.1}$& $<0.2$               & $10.9^{+5.6}_{-4.2}$& $<11.9$               & $<65.4$               & $<0.6$                & $<0.9$               & $1.9^{+1.1}_{-0.8}$& $<1.5$                & $10.2^{+12.4}_{-5.7}$  \\
\enddata
\tablecomments{$^a$ source name (NuSTAR~J); see Table~2; $^b$ origin
  of the X-ray data used in the spectral fitting; $^c$ observed-frame
  energy range (in keV) over which the X-ray data is fitted; $^d$
  best-fitting spectral slope ($\Gamma$) and uncertainty (90\%
  confidence) over the full spectral range for a power-law model; $^e$
  best-fitting spectral slope ($\Gamma$), uncertainty (90\%
  confidence), and luminosity (units of $10^{44}$~erg~s$^{-1}$) from
  fitting the rest-frame 10--40~keV data with a power-law model (see
  \S3.1 for more details); $^f$ best-fitting spectral slope
  ($\Gamma$), absorbing column density, uncertainty (90\% confidence),
  and upper limits ($N_{\rm H}$; units of $10^{22}$~cm$^{-2}$; see
  \S3.1 for more details).}
\end{deluxetable}
\clearpage
\end{landscape}

%
%

\clearpage
\begin{deluxetable}{lccc}
\tabletypesize{\small}
\tablewidth{0pt}
\tablecaption{Joint-fitting Model Parameters}
\tablehead{
\colhead{Model$^a$} &
\colhead{Sources$^b$} &
\colhead{$\Gamma_{\rm 10-40 keV}$$^c$} &
\colhead{$R$$^d$}}
\startdata
{\sc pow}    & 8 & $1.88^{+0.26}_{-0.25}$ & \nodata \\
{\sc pexrav} & 8 & 1.8$^e$ & $<1.4$ \\
{\sc pexrav} & 8 & $2.08^{+0.25}_{-0.24}$ & 1.0$^e$ \\
\enddata
\tablecomments{$^a$~{\sc xspec} model used in the joint-fitting
  process; $^b$~number of sources used in the joint-fitting process --
  the low-luminosity system NuSTAR~J032459-0256.1 and the HBL
  NuSTAR~J121027+3929.1 were not included in the joint-fitting
  process; $^c$~best-fitting spectral slope over the rest-frame
  10--40~keV range; $^d$~best-fitting reflection parameter ($R$; see
  Footnote 13 for a description) over the rest-frame 10--40~keV range;
  $^e$~parameter fixed at given value.}
\end{deluxetable}

\clearpage

\normalsize

%
%

\appendix

Here we provide the details of the new optical spectroscopy obtained
for eight of the serendipitous {\it NuSTAR} sources, present the
optical spectroscopy for an additional {\it Chandra}-detected source
in the Geminga field, and discuss the interesting properties of
NuSTAR~J183443+3237.8.\\

%
%
\figurenum{A1}
\begin{figure*}[!h]
{\includegraphics[angle=90,width=10.0cm]{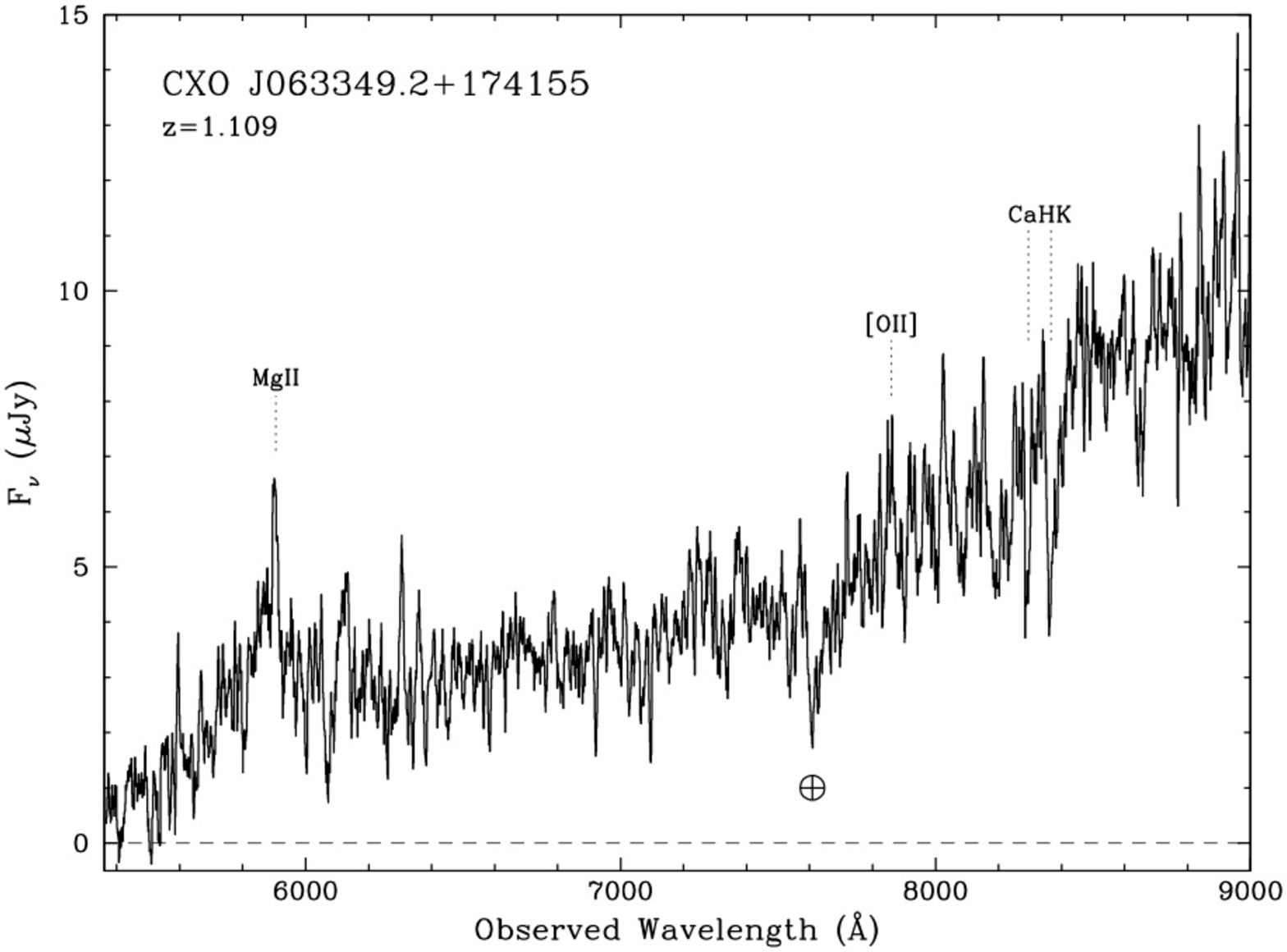}\hfill
  \includegraphics[angle=0,width=7.5cm]{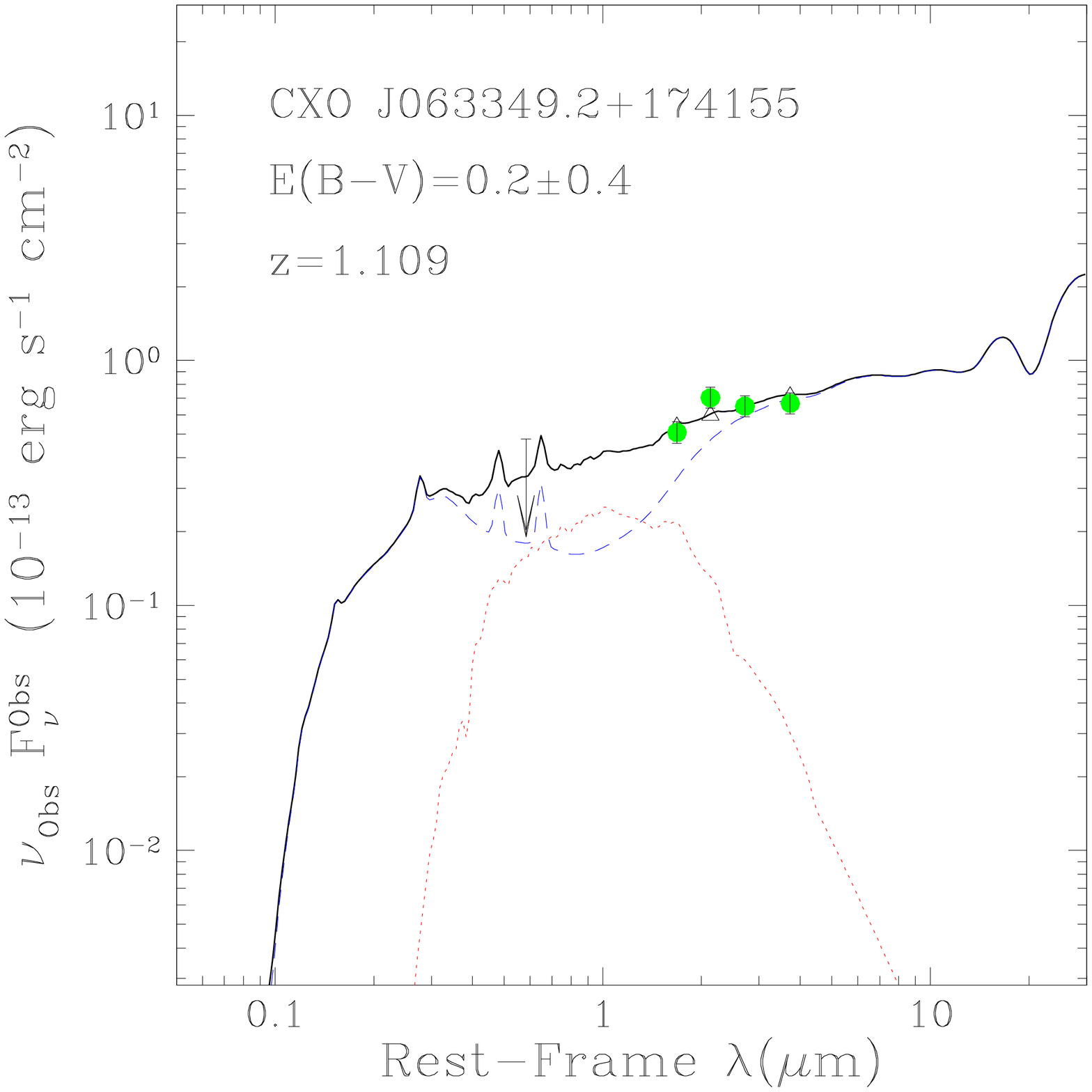}}
\caption{(left) optical spectrum of a {\it Chandra}-detected source
  spectroscopically identified in the Geminga field. The detection of
  broad Mg~II indicates that this source is a BLAGN at $z=1.109$;
  (right) UV--MIR SED with the best-fitting solution. The data are
  fitted with the Assef et~al. (2010) AGN (magenta dashed curve) and
  elliptical galaxy templates (red dotted curve). The best-fitting
  solution is plotted as a black solid curve. The source redshift,
  best-fitting dust-reddening solution ($E(B-V)$) and uncertainties
  are shown.}
\end{figure*}

\section{A.1 Details of the new optical spectrsocopic observations}

On UT 2012 October 10 we used the Double Spectrograph (DBSP) on the
Palomar 200~inch telescope to observe NuSTAR~J183443+3237.8 in the
3C382 field.  We integrated for 300~s split across two equal exposures
in moderate, but non-photometric conditions.  The observations used
the 2\farcs0 wide longslit, the 6800 \AA\ dichroic, the 600/4000 blue
grating (e.g., 600 $\ell$~${\rm mm}^{-1}$, blazed at 4000~\AA), and
the 316/7500 red grating.


On UT 2012 October 13 we used the DEep Imaging Multi-Object
Spectrograph (DEIMOS; Faber et~al. 2003) at the Nasymth focus of the
Keck~II 10~m telescope to observe NuSTAR~J204021-0056.1 in the AE Aqr
field.  We obtained a single 300~s exposure in photometric conditions
using 600/7500 grating.


On UT 2012 November 9 we used the Low Resolution Imaging Spectrometer
(LRIS; Oke et~al. 1995) at the Cassegrain focus of the Keck~I
telescope to observe NuSTAR~J032459-0256.1 and NuSTAR~J181428+3410.8 in
the NGC~1320 and W1814+3412 fields, respectively. We observed the
sources for 200~s and 300~s, respectively, in moderate, but
non-photometric conditions.  The observations used the 1\farcs5 wide
longslit, the 5600 \AA\ dichroic, the 400/3400 blue grism, and the
400/8500 red grating.


On UT 2012 November 20 we again used DBSP at Palomar. Conditions were
photometric and we used the same instrument configuration as employed
for NuSTAR~J183443+3237.8 in October. We observed
NuSTAR~J115746+6004.9 and NuSTAR~J115912+4232.6 in the SDSS~1157+6003
and IC~751 fields for 1800~s split into two and three dithered
exposures, respectively.


On UT 2012 December 12 we used the Gemini Multi-Object
Spectrograph-South (GMOS-S; Hook et~al. 2004) at the Gemini-South 8~m
telescope to observe NuSTAR~J011042-4604.2 in the HLX~1 field.  We
observed the source for 1200~s, split into two exposures dithered by
50~\AA\ in central wavelength to fill in the chip gap in the focal
plane.  We used the 1\farcs5 wide longslit and 600/4610 grating.


On UT 2013 January 10 we used LRIS at the Keck~I telescope to observe
NuSTAR~J063358+1742.4 in the Geminga field.  We observed the source for
1200~s, split into two exposures, using the 1\farcs5 wide longslit,
the 600/4000 blue grism, the 400/8500 red grating, and the
5600~\AA\ dichroic. The position angle of the longslit was set in
order to get a second {\it Chandra} source in the field where there is
weak evidence for {\it NuSTAR} emission. The optical spectrum of this
second Geminga serendipitous source at $\alpha_{\rm J2000} =
06^{h}33^m49.22^s$, $\delta_{\rm J2000} = +17\deg41\min55\farcs1$
(CXO~J063349.2+174155) and the UV--MIR SED and the best-fitting
solution (following \S3.2) are shown in Fig.~A1. The optical spectrum
reveals an AGN at $z = 1.109$ with somewhat broadened \ion{Mg}{2}
emission, weak [\ion{O}{2}] emission, and a strong 4000~\AA\ break
with well-detected Ca~H+K absorption lines. The best-fitting SED
solution suggests that the AGN dominates the UV--MIR
emission. However, since the SED is only comprised of the WISE data,
the overall SED is comparatively poorly constrained: the best-fitting
parameters are $\hat{a}=0.93\pm0.03$, $E(B-V)=0.24\pm0.39$, $L_{\rm
  6\mu m}=(5.34\pm0.55)\times10^{44}$~erg~s$^{-1}$, and
$M_{*}=(9.3\pm3.2)\times10^{10}$~$M_{\odot}$. There is weak evidence
for emission from this source in the {\it NuSTAR} images. However,
this source was not formally detected using the source detection
procedure described in \S2.1 and we therefore do not discuss this
source further in this paper. We instead provide this information for
future researchers of X-ray sources in the Geminga field.


%
%
\figurenum{A2}
\begin{figure*}[!h]
{\centerline{\includegraphics[angle=0,width=14.0cm]{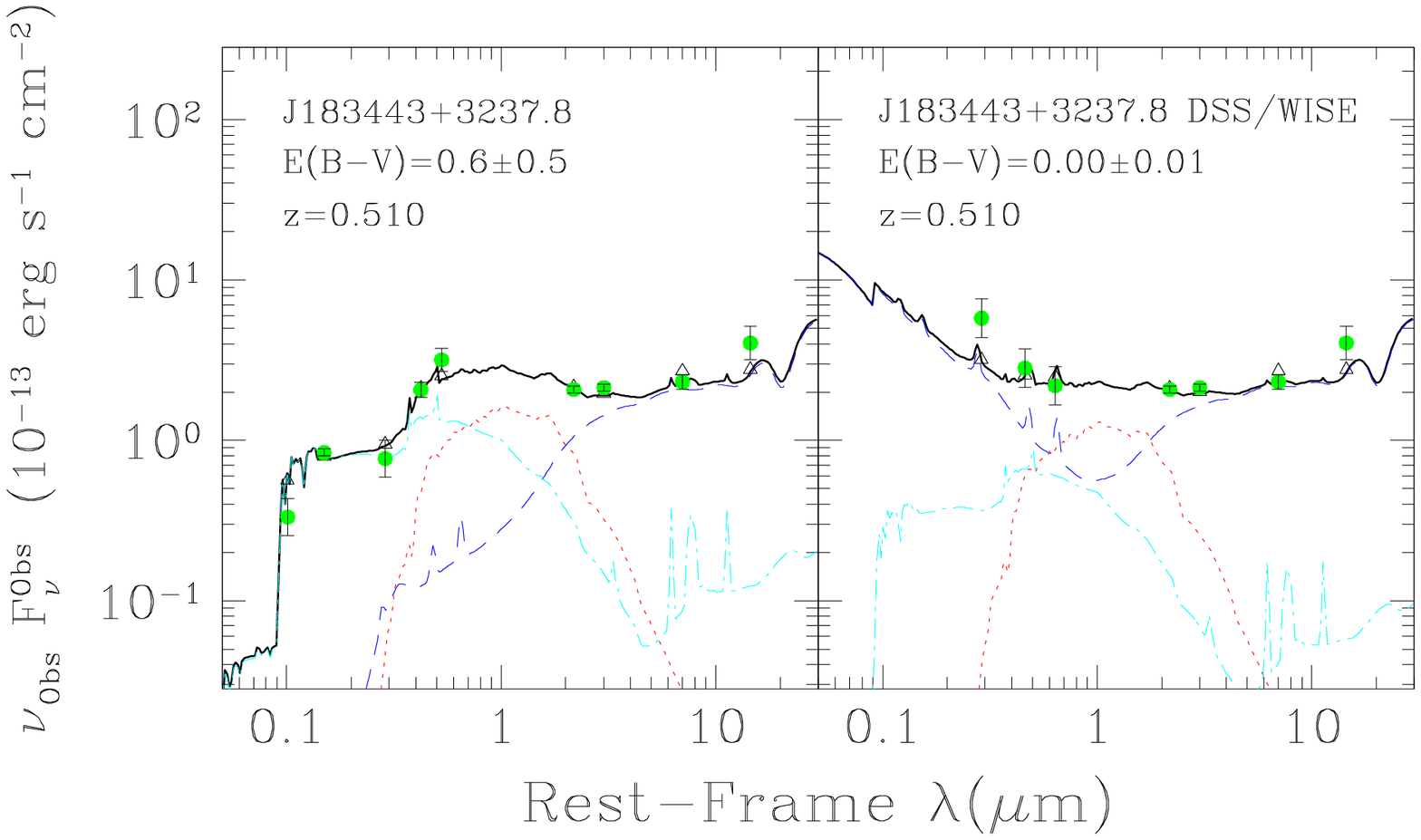}}}
\caption{UV--MIR SED and best-fitting solutions for
  NuSTAR~J183443+3237.8 using (left) our recent (UT 2013 March 04)
  observations and (right) from the DSS. The data are fitted with the
  Assef et~al. (2010) AGN (magenta dashed curve) and galaxy (elliptical: red dotted
  curve; irregular: cyan dash-dotted curve) templates; the best-fitting solution
  is plotted as a black solid curve. The source redshift, best-fitting
  dust-reddening solution ($E(B-V)$) and uncertainties are shown.}
\end{figure*}

\section{A.2 Notes on NuSTAR~J183443+32378}

NuSTAR~J183443+3237.8 is a BLAGN that appears to be unabsorbed in the
X-ray band ($N_{\rm H}<1.5\times10^{22}$~cm$^{-2}$; see Table~4). We
obtained $B$, $R$, and $I$ band observations of this field on UT 2013
March 04 using P60; see \S2.4. The optical emission of
NuSTAR~J183443+3237.8 has faded since the original DSS
observations. To explore the origin of this fading we fitted the
UV--MIR SED of NuSTAR~J183443+3237.8 following \S3.2, using both our
new data and the older DSS data; see Fig.~A2. On the basis of the
original DSS observations the best-fitting SED solution indicates
$E(B-V)=0.00\pm0.01$. However, by the comparison, the best-fitting SED
solution using the new UV--optical data indicates
$E(B-V)=0.59\pm0.46$, consistent with $A_{\rm V}\approx1.9$~mags for
$R_{\rm V}=3.1$ (e.g.,\ Savage \& Mathis 1979). Assuming the
relationship between dust reddening and X-ray absorption found in the
Galaxy (e.g.,\ G{\"u}ver \& {\"O}zel 2009), the X-ray absorbing column
density for $A_{\rm V}\approx1.9$~mags is $N_{\rm
  H}\approx5\times10^{21}$~cm$^{-2}$, a factor $\approx$~3 below the
upper limit placed on $N_{\rm H}$ from the X-ray spectral fitting; see
Table~4.

\end{document}